\documentclass[manuscript,screen]{acmart}

\usepackage{tikz}
\usepackage{array}
\usepackage{multirow}
\usetikzlibrary{arrows.meta,positioning,fit}

\newcommand{\rqanswer}[2]{%
  \par\addvspace{\baselineskip}
  \noindent\colorbox{gray!20}{%
    \parbox{\dimexpr\linewidth-2\fboxsep\relax}{%
      \textbf{Answer to #1.} #2}}
  \par}

\AtBeginDocument{%
  }

\setcopyright{none}
\begin{document}
\raggedbottom

\title{\texorpdfstring{HIDE: \underline{H}ierarchical \underline{I}solation via \underline{D}ocumentation and \underline{E}vidence for Rustc Bugs}{HIDE: Hierarchical Isolation via Documentation and Evidence for Rustc Bugs}}

\author{Tianrong Zhang}
\email{tianrongzhang25@stu.pku.edu.cn}
\affiliation{%
  \institution{Peking University Shenzhen Graduate School}
  \country{China}
}

\author{Yakun Zhang}
\email{zhangyk@hit.edu.cn}
\affiliation{%
  \institution{Harbin Institute of Technology, Shenzhen}
  \country{China}
}

\author{Dan Hao}
\email{haodan@pku.edu.cn}
\affiliation{%
  \institution{Peking University Shenzhen Graduate School}
  \country{China}
}

\renewcommand{\shortauthors}{Zhang et al.}

\begin{abstract}
Rustc bugs can prevent valid programs from compiling or cause incorrect compilation results, making the identification of responsible source files essential for repair.  Using large language models (LLMs) for rustc bug isolation raises three challenges.  First, distributed compiler functionality requires understanding both what individual files do and how they work together.  Second, a large source file search space makes exhaustive analysis of all files costly.  Third, including documentation and source code for many candidate files in LLM prompts leads to context overload, which can reduce bug isolation accuracy.  To address these challenges, we propose HIDE, a hierarchical LLM-based approach that combines compiler documentation with bug evidence to rank suspicious rustc source files.  Given a rustc revision and a bug context, HIDE consists of two stages that generate compiler documentation and narrow the search space.  To understand distributed compiler functionality, the \emph{Layered Documentation Aggregation} stage summarizes what individual files do and how files with similar functions work together.  To reduce the large source file search space, the \emph{Hierarchical Bug Isolation} stage uses hierarchical candidate selection to reduce the number of files requiring detailed analysis.  To mitigate context overload, this stage uses \emph{Scoped Candidate Competition} and \emph{Evidence-Guided Code Slicing} to limit the number of candidates and the amount of source code included in each prompt, respectively.  We evaluate HIDE on 160 reproducible rustc bugs and compare it with Agentless, SoapFL, and GraphLocator.  With GPT-5.4, HIDE achieves 42.08\% Accuracy@1 and 80.42\% Accuracy@10, exceeding the strongest baseline results by 5.83\% and 6.04\%, respectively.  These results demonstrate that generating compiler documentation, narrowing candidates hierarchically, and limiting analysis context enable effective rustc bug isolation.
\end{abstract}



\maketitle

\section{Introduction}
\label{sec:introduction}

Rustc is the primary compiler for Rust, a programming language that combines high execution efficiency with strong memory-safety and concurrency-safety guarantees~\cite{matsakis2014rust}.  Rustc translates source programs while performing checks and analyses to enforce these guarantees.  Rustc is therefore expected to compile programs correctly and enforce these guarantees so that Rust software can be built and trusted.  However, rustc bugs may terminate compilation, reject valid programs, or accept invalid programs~\cite{liu2025empirical}. These failures can block Rust software builds or undermine trust in the compiled software by producing incorrect compilation results.  To repair these bugs, developers first need to identify the responsible compiler source files.  This task, known as compiler bug isolation (CBI), is challenging in a large, evolving repository with strong dependencies among source files.  Effective compiler bug isolation can reduce the costs of understanding the compiler architecture, tracing compilation behavior, and isolating bugs.

Existing bug-isolation methods fall into two broad categories: \emph{general methods} and \emph{compiler-specific methods}.  General methods use semantic relevance, execution behavior, or code dependencies to locate faults and increasingly employ large language models (LLMs) to interpret such evidence~\cite{xia2025demystifying,li2025llm,yu2025orcaloca,kang2024quantitative,xu2025flexfl,qin2025s,chen2025locagent,jiang2025issue,liu2026graphlocator}.  Although broadly applicable, they do not explicitly exploit rustc's source organization and compiler execution behavior.  Compiler-specific methods obtain contrasting executions by mutating failing programs, varying compiler configurations, or skipping compilation steps~\cite{tu2024llm4cbi,yang2025odfl,liu2025isolating}.  Although LLM4CBI~\cite{tu2024llm4cbi} uses LLMs for tasks such as guiding program mutation, its localization evidence still comes primarily from execution differences, without reasoning about how dependent source files jointly implement compiler functionality.  LLM reasoning is therefore not unique to compiler bug isolation.  The open problem is how to combine it with rustc-specific structure and execution evidence.  This motivates using LLMs to understand compiler source code while exploiting rustc-specific information to guide bug isolation.

Using LLMs for rustc bug isolation raises three implementation challenges.  First, \textbf{distributed compiler functionality} requires understanding both what individual files do and how they work together.  Second, a \textbf{large source file search space} makes exhaustive analysis of all files costly.  Third, including documentation and source code for many candidate files in LLM prompts leads to \textbf{context overload}, which can reduce bug isolation accuracy.

To address these challenges, we propose HIDE, a hierarchical LLM-based approach that combines compiler documentation with bug evidence to rank suspicious rustc source files.  HIDE consists of two stages: \emph{Layered Documentation Aggregation} and \emph{Hierarchical Bug Isolation}.  Given a rustc revision and a bug context, HIDE generates compiler documentation to guide progressive candidate selection from crates to modules to files.  \textbf{To understand distributed compiler functionality,} the \emph{Layered Documentation Aggregation} stage summarizes what individual files do and how files with similar functions work together.  \textbf{To reduce the large source file search space,} the \emph{Hierarchical Bug Isolation} stage uses hierarchical candidate selection to reduce the number of files requiring detailed analysis.  \textbf{To mitigate context overload,} the \emph{Hierarchical Bug Isolation} stage uses two mechanisms, \emph{Scoped Candidate Competition} and \emph{Evidence-Guided Code Slicing}, to limit the number of candidates and the amount of source code included in each prompt, respectively.

We evaluate HIDE on 160 reproducible rustc bugs and compare it with Agentless~\cite{xia2025demystifying}, SoapFL~\cite{qin2025s}, and GraphLocator~\cite{liu2026graphlocator}, representing semantics-based, execution-based, and dependency-based approaches, respectively.  The evaluation yields four main findings.  \textbf{(1)} For effectiveness, HIDE achieves 42.08\% Accuracy@1 and 80.42\% Accuracy@10, exceeding the strongest baseline results by 5.83\% and 6.04\%, respectively.  \textbf{(2)} For efficiency, HIDE averages 154 seconds and 145,413 tokens per bug, achieving the lowest average execution time and consuming substantially fewer tokens than Agentless and SoapFL.  \textbf{(3)} For generalizability, HIDE obtains mean Accuracy@10 values between 76.04\% and 80.42\% with GPT-5.4, Gemini-3.1-Pro-Preview, and DeepSeek-V4-Pro, supporting its robustness across different LLMs.  \textbf{(4)} In the ablation study, all ablated variants achieve lower Accuracy@$k$ than HIDE, supporting the contributions of the proposed stages, components, and mechanisms.

Our main contributions are as follows:
\begin{itemize}
  \item We propose HIDE, a hierarchical LLM-based approach that combines compiler documentation with bug evidence to rank suspicious rustc source files.

  \item We propose the \emph{Layered Documentation Aggregation} stage, which provides hierarchical documentation summarizing what individual files do and how files with similar functions work together.

  \item We propose the \emph{Hierarchical Bug Isolation} stage, which uses hierarchical candidate selection to reduce the number of files requiring detailed analysis.  Its two mechanisms, \emph{Scoped Candidate Competition} and \emph{Evidence-Guided Code Slicing}, mitigate context overload by limiting the number of candidates and the amount of source code included in each prompt, respectively.

  \item We conduct a comprehensive evaluation on 160 reproducible rustc bugs with three LLMs and three representative baselines.  HIDE achieves the best results at every reported accuracy metric and the lowest average execution time.  Results across different LLMs support its robustness, while the ablation study supports the contributions of its stages, components, and mechanisms.  The code is available at \url{https://github.com/zhangtr02/HIDE}.
\end{itemize}

\section{Background}
\label{sec:background}

\subsection{Rustc Architecture and Query System}
\label{sec:rustc-architecture-query-system}

We identify two complementary sources of information in rustc that can support bug isolation: its static source organization and dynamic query execution.  The static source organization describes the directory structure of the compiler repository and provides candidate search scopes, while dynamic query execution is recorded in query traces that capture the computations involved during compilation.  We introduce both because combining code structure with execution evidence can help isolate rustc bugs.

\textbf{Static source organization.} Rustc organizes its source files hierarchically~\cite{rustcDevGuide}.  At the highest level, a \emph{crate} is a directory containing code for a set of compiler functions.  For example, the \texttt{rustc\_const\_eval} crate implements constant evaluation and MIR interpretation.  At the next level, a \emph{module} represented by a subdirectory groups files that implement specific compiler functionality.  For example, the \texttt{transform} module within this crate contains code for constant checking, constant promotion, and MIR validation.  At the finest level, a \emph{file} implements concrete checks, analyses, or transformations.  For example, the \texttt{validate.rs} file checks MIR structure, types, and storage usage.

\textbf{Dynamic query execution.} Rustc organizes compiler computations as \emph{queries}, each of which takes an input key and returns a result~\cite{rustcQuerySystem}.  For example, the \texttt{type\_of} query takes the identifier of a program item and returns its type.  A query is requested when its result is needed.  Rustc reuses a cached result when available and otherwise invokes the function implementing the query.  This function may request other queries, creating dependencies between computations.  A \emph{query trace} records query events observed during compilation, providing execution evidence for bug isolation.

\begin{figure}[t]
  \centering
  \begin{tikzpicture}[
  x=0.9cm,y=1cm,
  font=\sffamily\normalsize,
  >=Latex,
  panel/.style={draw=gray!55,rounded corners=4pt,fill=gray!2},
  crate/.style={draw=violet!65!black,rounded corners=3pt,fill=violet!5},
  module/.style={draw=blue!65!black,rounded corners=3pt,fill=blue!4},
  file/.style={draw=teal!65!black,rounded corners=2pt,fill=teal!5,
    minimum height=0.55cm},
  query/.style={draw=orange!75!black,rounded corners=2pt,fill=orange!6,
    minimum height=0.55cm,text width=3.4cm,align=center},
  card/.style={draw=gray!65,rounded corners=3pt,fill=white,
    minimum height=4.25cm,text width=4.2cm,align=left,inner sep=6pt},
  qflow/.style={->,semithick,draw=orange!80!black},
  cross/.style={->,dashed,semithick,draw=blue!65!black}
]
  \definecolor{bgblue}{HTML}{315E9B}
  \definecolor{bgred}{HTML}{A63737}

  \draw[panel] (0,12.65) rectangle (16.6,6.25);
  \node[anchor=west,font=\bfseries\normalsize] at (0.25,12.33)
    {(a) Information for rustc bug isolation};

  \node[anchor=west,font=\bfseries,text=violet!70!black]
    at (0.45,11.78) {Static source organization};
  \node[crate,minimum width=6.3cm,minimum height=4.6cm,anchor=north west]
    (crate) at (0.45,11.37) {};
  \node[anchor=west,font=\bfseries,text=violet!70!black]
    at (0.72,11.07) {\texttt{rustc\_const\_eval} crate};
  \node[anchor=west,font=\normalsize] at (0.72,10.67)
    {Constant evaluation and MIR interpretation};
  \node[module,minimum width=5.55cm,minimum height=3.1cm,anchor=north west]
    (module) at (0.98,10.37) {};
  \node[anchor=west,font=\bfseries,text=blue!70!black]
    at (1.24,10.07) {\texttt{transform} module};
  \node[file,text width=4.8cm,align=center] (validate) at (4.05,9.25)
    {\texttt{validate.rs} file\\MIR validation};
  \node[file,text width=4.8cm,align=center] (promote) at (4.05,8.25)
    {\texttt{promote\_consts.rs} file\\Constant promotion};
  \node[font=\normalsize,text=gray!70!black] at (3.95,7.65)
    {Other files in this module};
  \node[font=\normalsize,text=gray!70!black] at (3.95,7.0)
    {Nested boxes show directory containment};

  \node[anchor=west,font=\bfseries,text=orange!75!black]
    at (8.15,11.78) {Dynamic query execution};
  \node[anchor=west,font=\normalsize,text=orange!75!black]
    at (8.15,11.35) {Orange arrow $A \rightarrow B$ means $A$ requests $B$};
  \node[query] (promoted) at (12.10,10.05) {\texttt{mir\_promoted}};
  \node[query,text width=3.1cm] (qualif) at (9.95,8.75)
    {\texttt{mir\_const\_qualif}};
  \node[query,text width=3.1cm] (constant) at (14.05,8.75)
    {\texttt{mir\_const}};
  \node[query,text width=3.1cm] (built) at (14.05,7.55)
    {\texttt{mir\_built}};
  \draw[qflow] (promoted.south west) -- (qualif.north);
  \draw[qflow] (promoted.south east) -- (constant.north);
  \draw[qflow] (qualif.east) -- (constant.west);
  \draw[qflow] (constant.south) -- (built.north);
  \node[font=\normalsize,text=gray!70!black,align=center] at (10.1,7.55)
    {Example query requests\\when compiling a constant};
  \draw[cross,rounded corners=4pt] (promoted.west) -- (7.7,10.05)
    -- (7.7,8.25) -- (promote.east);
  \node[font=\normalsize,text=blue!65!black,align=center,fill=gray!2,inner sep=2pt]
    at (8.85,10.05) {Implementation\\code used\\by the query};
  \node[font=\normalsize,text=gray!80!black] at (8.3,6.53)
    {Execution evidence guides candidate selection within directory scopes};
  \draw[panel] (0,5.85) rectangle (16.6,-1.2);
  \node[anchor=west,font=\bfseries\normalsize] at (0.25,5.52)
    {(b) Motivating Bug 118990};

  \node[card,anchor=north west] (program) at (0.35,5.05) {%
    \begin{minipage}[t][3.8cm][t]{\linewidth}
    \centering\textbf{Failure Program}\par\vspace{3pt}
    \raggedright
    {\ttfamily\normalsize
      \setlength{\baselineskip}{10pt}%
      struct CompileError;\\
      impl Drop for\\
      CompileError \{...\}\\[2pt]
      'code: \{\\
      \ \ break 'code;\\
      \ \ CompileError.throw();\\
      \}\par}
    \vspace{4pt}
    \textcolor{bgblue}{Early exit skips temporary creation.}
    \end{minipage}
  };
  \node[card,anchor=north west] (ice) at (5.85,5.05) {%
    \begin{minipage}[t][3.8cm][t]{\linewidth}
    \centering\textbf{Validation Error}\par\vspace{5pt}
    \raggedright
    \textbf{Reported error}\par
    \textcolor{bgred}{Cleanup of a temporary value whose storage is not active.}\par\vspace{6pt}
    \textbf{Why the report is incorrect}\par
    Cleanup would not occur after the early exit.
    \end{minipage}
  };
  \node[card,anchor=north west] (trace) at (11.35,5.05) {%
    \begin{minipage}[t][3.8cm][t]{\linewidth}
    \centering\textbf{Query Trace}\par\vspace{4pt}
    \raggedright
    \texttt{mir\_built} --- MIR construction\\[3pt]
    \texttt{mir\_borrowck} --- borrow checking\\
    \ldots\par\vspace{5pt}
    These tasks also occur during normal compilation. Their presence does not identify the responsible files.
    \end{minipage}
  };
  \node[anchor=west,font=\bfseries] at (0.35,0.42)
    {Validation across files in different crates};
  \node[file,text width=4.0cm,minimum height=1.1cm,align=center]
    (storage) at (2.88,-0.4)
    {\texttt{storage\_liveness.rs}\\Provides storage information};
  \node[file,text width=4.0cm,minimum height=1.1cm,align=center]
    (bugvalidate) at (8.38,-0.4)
    {\texttt{validate.rs}\\Checks MIR using storage information};
  \node[file,text width=4.0cm,minimum height=1.1cm,align=center]
    (manager) at (13.88,-0.4)
    {\texttt{pass\_manager.rs}\\Controls when validation runs};
  \draw[->,semithick,draw=teal!70!black] (storage.east) -- (bugvalidate.west);
  \draw[->,semithick,draw=teal!70!black] (manager.west) -- (bugvalidate.east);

\end{tikzpicture}
  \caption{Static and dynamic evidence for rustc bug isolation and motivating Bug 118990.}
  \Description{Panel (a) has two sides labeled static source organization and dynamic query execution. Nested boxes on the left show the rustc_const_eval crate, its transform module, and the validate.rs and promote_consts.rs files. On the right, orange arrows show mir_promoted requesting mir_const_qualif and mir_const, mir_const_qualif requesting mir_const, and mir_const requesting mir_built. A dashed arrow points from mir_promoted to promote_consts.rs to indicate code used during that query. A bottom annotation states that execution evidence guides candidate selection within directory scopes. Panel (b) shows the failure program, an explanation of the incorrect validation report, and a shortened query trace. An early exit skips temporary creation, so cleanup would not occur, but the check overlooks this condition. The query trace includes MIR construction and borrow checking, which also occur during normal compilation. Below, arrows from storage_liveness.rs and pass_manager.rs to validate.rs indicate storage information supplied to the check and control over when validation runs. These files belong to different crates.}
  \label{fig:background}
\end{figure}
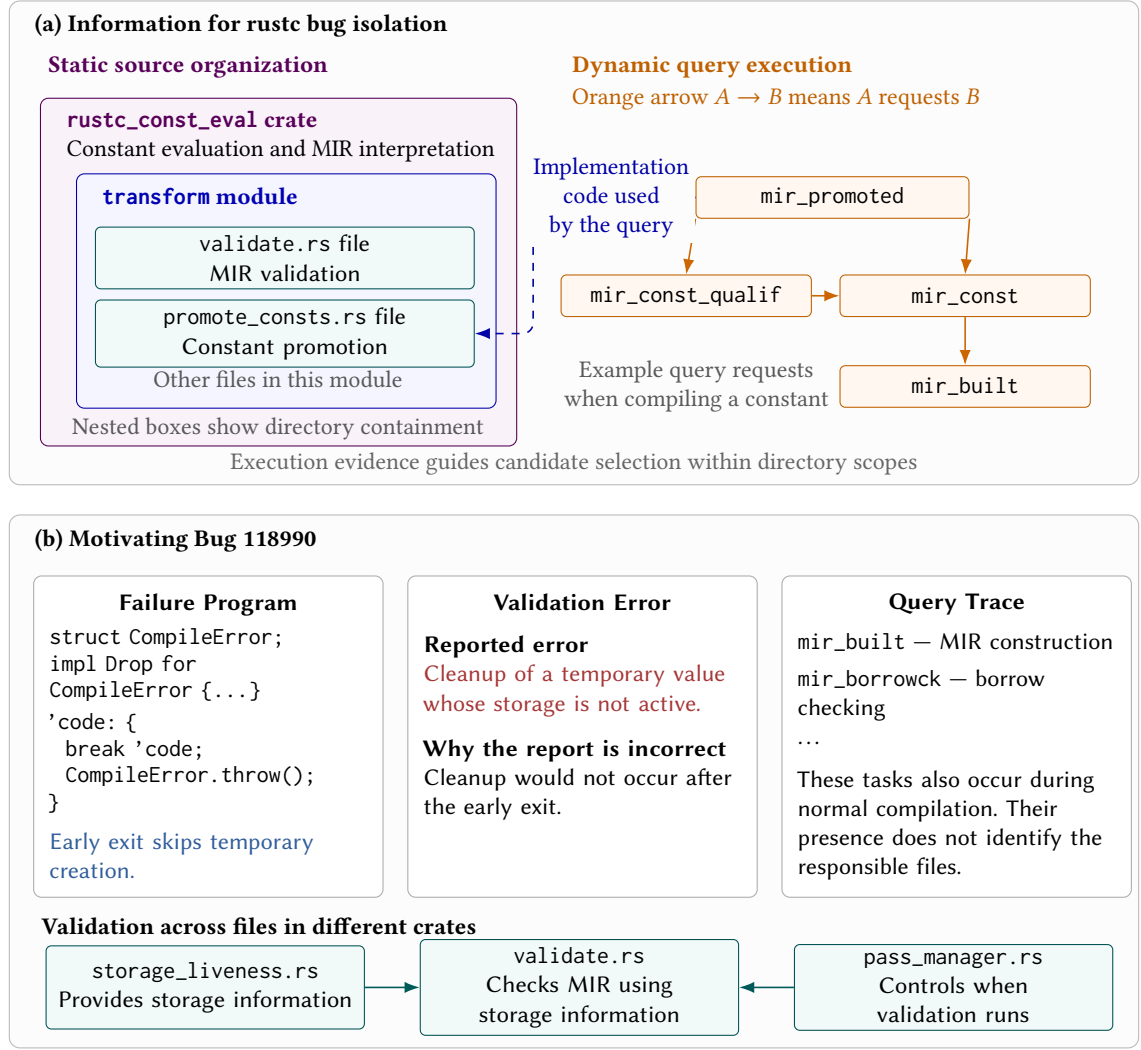

Taken together, static source organization and dynamic query execution support bug isolation in complementary ways.  The former divides compiler files into candidate search scopes, while the latter provides evidence of the computations performed during compilation.  They allow us to use execution evidence to guide the search for suspicious files within these scopes.  Figure~\ref{fig:background}(a) illustrates this combination.  The nested boxes show the containment of crates, modules, and files.  The query nodes and orange arrows illustrate selected query requests when compiling a constant body.  The dashed arrow points from the \texttt{mir\_promoted} query to the \texttt{promote\_consts.rs} file, whose constant promotion code is used during this query.

\subsection{Motivating Example}
\label{sec:motivating-example}

We use Bug 118990~\cite{rustcIssue118990} to illustrate the challenges of rustc bug isolation.  Following Section~\ref{sec:rustc-architecture-query-system}, we examine static source organization and dynamic query execution alongside the bug report, compiler log, and failure program, which is the program that triggers the compiler failure.  Figure~\ref{fig:background}(b) shows this program, explains the incorrect report, summarizes the query trace and the files involved in validation.  In the failure program, \texttt{break} exits a block before a temporary \texttt{CompileError} value is created.  This type uses \texttt{Drop} to define cleanup when a value is destroyed.  With \texttt{-Zvalidate-mir} enabled, rustc checks its mid-level intermediate representation (MIR), an internal form of the program, and terminates with an internal compiler error (ICE).  The check reports a cleanup operation on a temporary value whose storage is not active.  Since cleanup at this stage is conditional on initialization, the early exit means that this cleanup would not occur.  The check overlooks this condition and therefore incorrectly reports an error~\cite{rustcPR119077}.

First, Bug 118990 illustrates the challenge of understanding \textbf{distributed compiler functionality}.  Explaining the incorrect report requires understanding both the validation rule and the storage information it relies on.  The \texttt{validate.rs} file in the \texttt{rustc\_const\_eval} crate checks MIR using information about when storage is active, provided by the \texttt{storage\_liveness.rs} file in the \texttt{rustc\_mir\_dataflow} crate.  The \texttt{pass\_manager.rs} file in the \texttt{rustc\_mir\_transform} crate controls when validation runs~\cite{rustcPR119077}.  An LLM must understand how these files work together to judge whether the check accounts for the condition under which cleanup occurs.  General bug-isolation methods identify candidate files based on semantic relevance between problem descriptions and code.  In this example, isolating the bug further requires understanding how storage information and the timing of validation affect the check.

Second, Bug 118990 illustrates the challenge of a \textbf{large source file search space}.  The query trace in Figure~\ref{fig:background}(b) includes \texttt{mir\_built} and \texttt{mir\_borrowck}, corresponding to MIR construction and borrow checking.  These tasks also occur during normal compilation, so their presence alone does not tell us which files caused the failure.  Following the code that implements these tasks and its dependencies can lead to many files in different crates and modules.  Static source organization groups these files into search scopes, but we still need to decide which scopes and files to inspect.  General bug-isolation methods use execution information and code dependencies to reduce the search space.  In this example, an LLM must still determine which of the files involved in compilation could be responsible for the bug.  Examining every candidate file in detail is costly, making it necessary to narrow the candidates before detailed analysis.

Finally, Bug 118990 illustrates the challenge of \textbf{context overload}.  An LLM analyzing this bug needs to understand the validation rule, the storage information it relies on, and when validation runs.  Including many candidate files in one prompt requires the LLM to assess how each file relates to the failure.  Providing complete source code for each candidate also requires it to read code unrelated to the failure.  These two demands can make it harder to determine whether cleanup would actually occur.  General bug-isolation methods face this difficulty when they include documentation and source code for many candidate files in the same prompt.  This example motivates comparing smaller groups of candidate files and focusing on bug-related code within each file.

Overall, Bug 118990 illustrates three challenges of rustc bug isolation, namely distributed compiler functionality, a large source file search space, and context overload.  Addressing these challenges requires understanding what individual files do and how they work together, narrowing candidates before detailed analysis, and comparing smaller groups of files using bug-related code.  These requirements motivate Rustc-CBI, a hierarchical LLM-based approach to rustc bug isolation, which we present in the next section.

\section{Approach}
\label{sec:approach}

\begin{figure}[t]
  \centering
  \includegraphics[width=\linewidth]{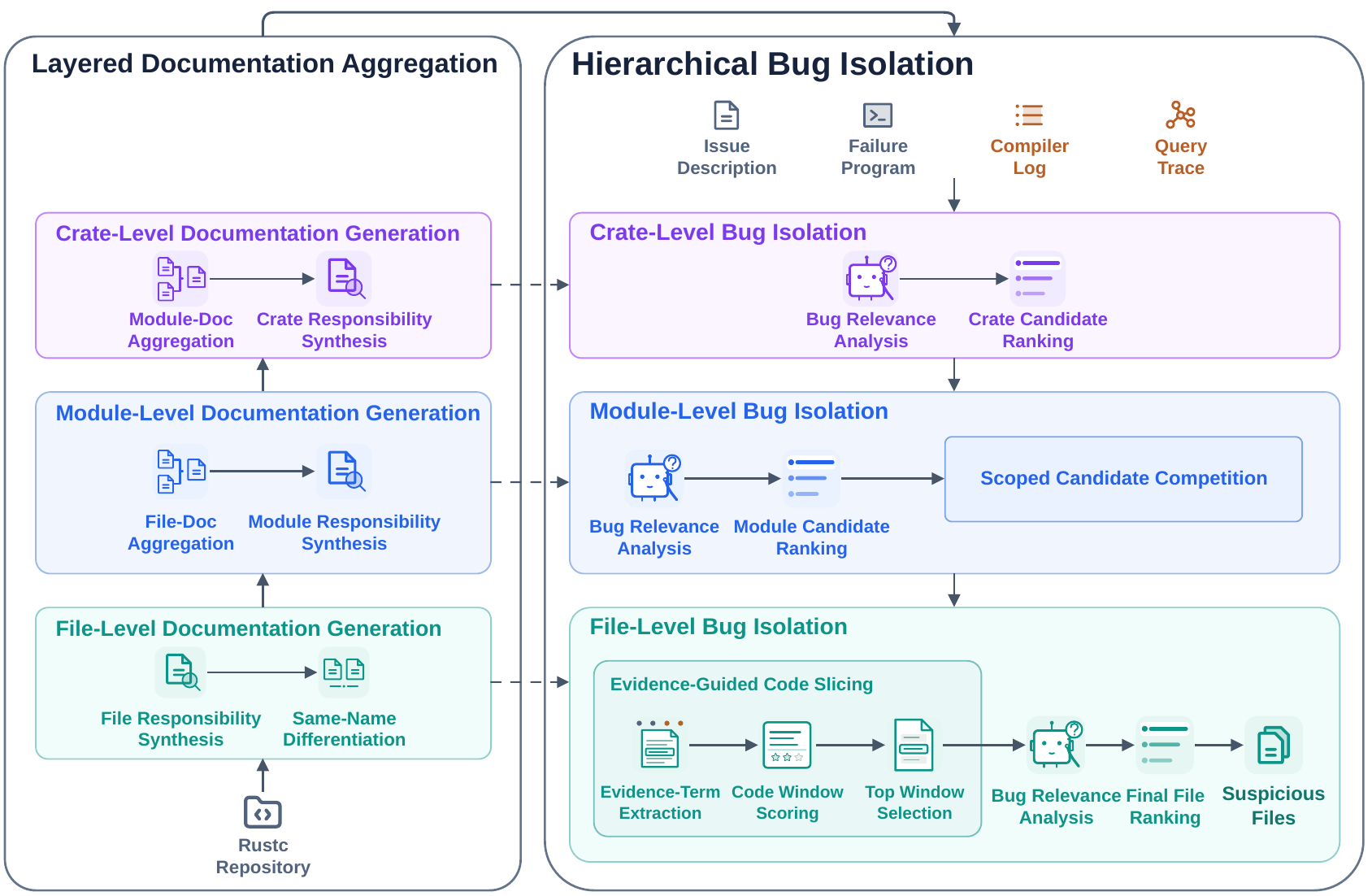}
  \caption{Overview of HIDE.}
  \Description{HIDE is a hierarchical LLM-based approach that combines compiler documentation with bug evidence to rank suspicious rustc source files. It takes a rustc revision and a bug context as inputs. The layered documentation aggregation stage contains the file-level documentation generation, module-level documentation generation, and crate-level documentation generation components, which operate bottom-up. The hierarchical bug isolation stage contains the crate-level bug isolation, module-level bug isolation, and file-level bug isolation components, which operate top-down using a shared bug context containing the issue description, failure program, compiler log, and query trace. The scoped candidate competition mechanism operates within the module-level bug isolation component to compare files within individual modules, limiting the number of candidates considered together. The evidence-guided code slicing mechanism operates within the file-level bug isolation component to provide bug-related code slices instead of complete files, limiting the amount of source code examined.}
  \label{fig:approach-overview}
\end{figure}

Figure~\ref{fig:approach-overview} presents an overview of HIDE.  HIDE is a hierarchical LLM-based approach that combines compiler documentation with bug evidence to rank suspicious rustc source files.  Given a rustc revision and a bug context, it progressively narrows candidates from crates to modules to files.  HIDE consists of two stages.  The \emph{Layered Documentation Aggregation} stage contains the \emph{File-Level Documentation Generation}, \emph{Module-Level Documentation Generation}, and \emph{Crate-Level Documentation Generation} components.  The \emph{Hierarchical Bug Isolation} stage contains the \emph{Crate-Level Bug Isolation}, \emph{Module-Level Bug Isolation}, and \emph{File-Level Bug Isolation} components.

The \emph{Layered Documentation Aggregation} stage uses LLMs to describe compiler functionality at the file, module, and crate levels.  Its three components operate from files to modules and then to crates.  The file-level component generates a file doc from each file's path and source code, the module-level component aggregates file docs within each module, and the crate-level component aggregates module docs within each crate.  Together, these docs summarize file functionality and relationships across modules to guide hierarchical search.

The \emph{Hierarchical Bug Isolation} stage uses the docs and bug context to progressively narrow candidates from crates to modules and then to files.  At each level, the corresponding component assesses the current candidates, selects a smaller set, and passes its assessments to the next component.  Within this stage, the \emph{Scoped Candidate Competition} mechanism compares files separately within each selected module before global ranking, limiting the number of files considered together.  The \emph{Evidence-Guided Code Slicing} mechanism supplies code slices selected using the bug context instead of complete files, limiting the amount of source code examined.

\subsection{Layered Documentation Aggregation}
\label{sec:documentation-aggregation}

The \emph{Layered Documentation Aggregation} stage summarizes what individual files do and how files with similar functions work together.  Given a rustc revision, its three components construct compiler documentation from files to modules and then to crates.  \emph{File-Level Documentation Generation} component summarizes each file's responsibilities from its repository path and source code, producing file docs $\mathcal{D}_f$.  \emph{Module-Level Documentation Generation} component aggregates file docs to describe each module's functionality and how its files work together, producing module docs $\mathcal{D}_m$.  \emph{Crate-Level Documentation Generation} component aggregates module docs to describe each crate's functionality and how its modules work together, producing crate docs $\mathcal{D}_c$.  These file, module, and crate docs serve as inputs to the \emph{Hierarchical Bug Isolation} stage, which uses them to select candidates at the corresponding levels, as described in Section~\ref{sec:hierarchical-isolation}.  Below, we first describe the three components and then present the prompt design that implements them through a shared template with level-specific inputs and output requirements.

The first component, \emph{File-Level Documentation Generation}, produces a file doc $d_f(f)$ that summarizes the compiler functionality implemented by each Rust source file $f$.  It first performs \emph{File Responsibility Synthesis}, using the file's repository path and source code to summarize the checks, analyses, or transformations it implements without focusing on a particular bug.  It then performs \emph{Same-Name Differentiation}, using the file's repository path and source code to clarify functional differences between files with identical names in different modules or crates.  The refined description becomes the file doc.  For example, the file doc for the \texttt{validate.rs} file in the \texttt{rustc\_const\_eval} crate describes its checks of MIR structure, types, and storage usage.

The second component, \emph{Module-Level Documentation Generation}, produces a module doc $d_m(m)$ that summarizes the compiler functionality implemented by each module $m$ and how its files work together.  It first performs \emph{File-Doc Aggregation}, collecting the file docs within the module according to the repository's directory structure.  It then performs \emph{Module Responsibility Synthesis}, using these file docs to summarize the module's compiler functionality and how its files work together.  The resulting module doc provides a combined summary without repeating every file doc.  For example, the module doc for the \texttt{transform} module in the \texttt{rustc\_const\_eval} crate describes constant checking, constant promotion, and MIR validation.

The third component, \emph{Crate-Level Documentation Generation}, produces a crate doc $d_c(c)$ that summarizes the compiler functionality implemented by each crate $c$ and how its modules work together.  It first performs \emph{Module-Doc Aggregation}, collecting the module docs within the crate according to the repository's directory structure.  It then performs \emph{Crate Responsibility Synthesis}, using these module docs to summarize the crate's compiler functionality and how its modules work together.  The resulting crate doc provides a combined summary without repeating every module doc.  For example, the crate doc for the \texttt{rustc\_const\_eval} crate describes constant evaluation, MIR interpretation, and MIR validation.  We express the generation of module docs and crate docs as follows.
\begin{align}
  d_m(m) &=
  \mathcal{A}\bigl(\{d_f(f) \mid f \in m\}\bigr),\\
  d_c(c) &=
  \mathcal{A}\bigl(\{d_m(m) \mid m \in c\}\bigr).
\end{align}
Here, $\mathcal{A}$ denotes LLM-based document aggregation, which summarizes compiler functionality from the input docs.  A module doc is generated from the file docs within that module, and a crate doc from the module docs within that crate.

\paragraph{Prompt Design.}
The three components use the shared prompt template in Figure~\ref{fig:documentation-prompt}, with their inputs, processing, and outputs summarized in Table~\ref{tab:documentation-prompts}.  The template identifies the target file, module, or crate and specifies its level and input content.  At the file level, the input contains the repository path and source code, and the prompt asks for the checks, analyses, or transformations implemented by the file and any necessary distinctions between files with identical names.  At the module and crate levels, the inputs contain file docs and module docs, respectively, and the prompt asks for the compiler functionality they collectively describe and how the files or modules work together.  For each component, the LLM follows the two steps listed in Table~\ref{tab:documentation-prompts} to generate the corresponding doc.

\begin{figure}[t]
  \centering
  \begin{tikzpicture}[x=1cm,y=1cm]
  \definecolor{promptborder}{HTML}{5B6F8A}
  \definecolor{promptfill}{HTML}{F7F9FC}
  \definecolor{prompthead}{HTML}{E8EEF7}
  \definecolor{promptlabel}{HTML}{315E9B}

  \node[anchor=north west,draw=promptborder,rounded corners=4pt,
        line width=0.8pt,fill=promptfill,inner sep=0pt,text=black]
    at (0,0) {%
    \begin{minipage}{14.8cm}
      \colorbox{prompthead}{%
        \parbox[c][0.6cm][c]{\dimexpr\linewidth-2\fboxsep\relax}{%
          \centering\sffamily\bfseries\normalsize\color{promptborder}
          Prompt Template}}\par
      \vspace{6pt}
      \noindent\hspace*{0.4cm}%
      \begin{minipage}{14.0cm}
      \ttfamily\normalsize\raggedright
      You are analyzing the architecture of the rustc compiler.
      \\[0.55em]
      \textcolor{promptlabel}{Target level:}
      \textless level\textgreater\\
      \textcolor{promptlabel}{Target entity:}
      \textless entity\_id\textgreater\\
      \textcolor{promptlabel}{Input content:}
      \textless input\_content\textgreater
      \\[0.55em]
      \textcolor{promptlabel}{Instructions:}\\
      Summarize the compiler functionality implemented by the target file, module, or crate. For a file, use its repository path and source code to describe its checks, analyses, or transformations and clarify functional differences between files with identical names. For a module or crate, use the input docs to summarize its compiler functionality and how its files or modules work together.
      \\[0.55em]
      \textcolor{promptlabel}{Output:}\\
      Return a concise structured description.
      \end{minipage}\par
      \vspace{8pt}
    \end{minipage}};
\end{tikzpicture}
  \caption{Prompt template for \emph{Layered Documentation Aggregation}.}
  \Description{A framed prompt template presents a task statement, input fields, instructions, and an output requirement. The fields specify the target level, target entity, and input content. The instructions request a summary of compiler functionality from source code for a file or from input docs for a module or crate, including distinctions between files with identical names and how files or modules work together. The output is a concise structured description.}
  \label{fig:documentation-prompt}
  \par\medskip
  \begin{minipage}{\linewidth}
  \captionof{table}{Prompt design across the three documentation components.}
  \label{tab:documentation-prompts}
  \centering
  \small
  \setlength{\tabcolsep}{2pt}
  \renewcommand{\arraystretch}{1.15}
  \begin{tabular}{@{}>{\raggedright\arraybackslash}p{0.31\linewidth}
                      >{\raggedright\arraybackslash}p{0.28\linewidth}
                      >{\raggedright\arraybackslash}p{0.275\linewidth}
                      >{\raggedright\arraybackslash}p{0.09\linewidth}@{}}
    \toprule
    Component & Input content & Processing & Output \\
    \midrule
    \emph{File-Level Documentation Generation}
      & Repository path and source code
      & \emph{File Responsibility Synthesis}\par
        \emph{Same-Name Differentiation}
      & File doc \\
    \midrule
    \emph{Module-Level Documentation Generation}
      & File docs within the target module
      & \emph{File-Doc Aggregation}\par
        \emph{Module Responsibility Synthesis}
      & Module doc \\
    \midrule
    \emph{Crate-Level Documentation Generation}
      & Module docs within the target crate
      & \emph{Module-Doc Aggregation}\par
        \emph{Crate Responsibility Synthesis}
      & Crate doc \\
    \bottomrule
  \end{tabular}
  \end{minipage}
\end{figure}

In summary, the \emph{Layered Documentation Aggregation} stage provides docs that describe compiler functionality at the file, module, and crate levels, including how files and modules work together.  The \emph{Hierarchical Bug Isolation} stage uses these docs with the bug context to select candidates at each level.

\subsection{Hierarchical Bug Isolation}
\label{sec:hierarchical-isolation}

The \emph{Hierarchical Bug Isolation} stage uses hierarchical candidate selection to reduce the number of files requiring detailed analysis.  It combines the docs from the \emph{Layered Documentation Aggregation} stage with a shared bug context to progressively narrow candidates from crates to modules and then to files, producing a ranked list of suspicious files.  To mitigate context overload, it uses two mechanisms.  At the module level, the \emph{Scoped Candidate Competition} mechanism compares files within individual modules before global ranking, limiting the number of candidates considered together.  At the file level, the \emph{Evidence-Guided Code Slicing} mechanism provides bug-related code slices instead of complete files, limiting the amount of source code examined.  This stage consists of three components, \emph{Crate-Level Bug Isolation}, \emph{Module-Level Bug Isolation}, and \emph{File-Level Bug Isolation}, which we describe below after defining the shared bug context.  We then present the prompt design shared by the three components and the \emph{Scoped Candidate Competition} mechanism.

All three components use the same bug context, shown on the right side of Figure~\ref{fig:approach-overview}.  We represent this context as
\begin{equation}
  \mathcal{B}=\langle I,P,L,Q\rangle,
\end{equation}
where $I$, $P$, $L$, and $Q$ denote the issue description, failure program, compiler log, and query trace, respectively.  The issue description and failure program describe the failure and its triggering conditions, while the compiler log and query trace provide evidence of compilation behavior and query execution.  At each level, the corresponding component uses this bug context with the docs at that level to select candidates.  The crate-level and module-level components pass the candidates and the reasons for their selection to the next component to guide further analysis.

\subsubsection{Crate-Level Bug Isolation}
\label{sec:crate-isolation}

The first component, \emph{Crate-Level Bug Isolation}, selects crate candidates for the \emph{Module-Level Bug Isolation} component.  It collects all compiler crates in the target rustc revision to form the initial candidate set $\mathcal{C}$.  It first performs \emph{Crate-Level Bug Relevance Analysis}, using the crate docs and bug context to assess whether each crate may be responsible for the bug.  It then performs \emph{Crate Candidate Ranking}, comparing these assessments to rank the crates and select a small set of candidates.

During \emph{Crate-Level Bug Relevance Analysis}, the LLM analyzes each crate $c\in\mathcal{C}$ using its crate doc $d_c(c)$ and the bug context $\mathcal{B}$.  Based on the bug context, it assesses whether the compiler functionality described in the crate doc may be responsible for the bug.  For each crate, the LLM produces an assessment for use in \emph{Crate Candidate Ranking}.

During \emph{Crate Candidate Ranking}, the LLM compares the crate-level assessments to rank all crates in $\mathcal{C}$.  Crates considered more likely to be responsible for the bug are ranked higher.  It selects the top-$K_c$ crates as the candidate set
\begin{equation}
  \mathcal{S}_c=
  \operatorname{TopK}_{K_c}(\mathcal{C}),
\end{equation}
where $K_c$ is the number of crate candidates, and $\operatorname{TopK}_{K_c}$ selects the $K_c$ highest-ranked crates based on these assessments.  The \emph{Module-Level Bug Isolation} component then searches only the modules within these crate candidates.

\subsubsection{Module-Level Bug Isolation}
\label{sec:module-isolation}

The second component, \emph{Module-Level Bug Isolation}, selects module candidates within the crate candidates.  It collects all modules in the crate candidates $\mathcal{S}_c$ to form the initial candidate set $\mathcal{M}$.  It first performs \emph{Module-Level Bug Relevance Analysis}, using the module docs, bug context, and crate-level assessments to assess whether each module may be responsible for the bug.  It then performs \emph{Module Candidate Ranking}, comparing these assessments to rank the modules and select a small set of candidates.  The \emph{Scoped Candidate Competition} mechanism subsequently selects files separately within each module candidate.

During \emph{Module-Level Bug Relevance Analysis}, the LLM analyzes each module $m\in\mathcal{M}$ using its module doc $d_m(m)$, the bug context $\mathcal{B}$, and the assessment of its owning crate.  Based on the bug context and crate-level assessment, it assesses whether the compiler functionality described in the module doc may be responsible for the bug.  For each module, the LLM produces an assessment for use in \emph{Module Candidate Ranking}.

During \emph{Module Candidate Ranking}, the LLM compares the module-level assessments to rank all modules in $\mathcal{M}$.  Modules considered more likely to be responsible for the bug are ranked higher.  It selects the top-$K_m$ modules as the candidate set
\begin{equation}
  \mathcal{S}_m=
  \operatorname{TopK}_{K_m}(\mathcal{M}),
\end{equation}
where $K_m$ is the number of module candidates, and $\operatorname{TopK}_{K_m}$ selects the $K_m$ highest-ranked modules based on these assessments.  This comparison ranks modules across all crate candidates without requiring each crate to contribute the same number of modules.  The \emph{Scoped Candidate Competition} mechanism then selects files separately within each module candidate, using the module-level assessments to guide the analysis.

\begin{figure}[t]
  \centering
  \begin{tikzpicture}[
  font=\sffamily\normalsize,
  >={Latex[length=2mm]},
  arr/.style={->,draw=gray!75,line width=0.7pt},
  mod/.style={draw=gray!65,rounded corners=2pt,fill=white,
    text width=2.45cm,minimum height=0.78cm,align=center},
  scope/.style={draw=blue!60,rounded corners=2pt,fill=blue!3,
    text width=2.65cm,minimum height=0.82cm,align=center},
  pool/.style={draw=gray!70,rounded corners=2pt,fill=white,
    text width=2.25cm,minimum height=2.05cm,align=center},
  combined/.style={draw=blue!70,rounded corners=2pt,fill=white,
    text width=2.45cm,minimum height=2.05cm,align=center}
]
  \draw[rounded corners=5pt,draw=gray!60,fill=gray!3]
    (-6.65,-2.4) rectangle (-0.2,2.85);
  \draw[rounded corners=5pt,draw=blue!60,fill=blue!4]
    (0.2,-2.4) rectangle (6.65,2.85);
  \node[font=\sffamily\normalsize\bfseries] at (-3.43,2.45)
    {Naive Global Competition};
  \node[font=\sffamily\normalsize\bfseries,text=blue!75] at (3.43,2.45)
    {Scoped Candidate Competition};

  \node[mod] (la) at (-5.0,1.4) {\textbf{Module A}\\18 files};
  \node[mod] (lb) at (-5.0,0) {\textbf{Module B}\\7 files};
  \node[mod] (lc) at (-5.0,-1.4) {\textbf{Module C}\\3 files};
  \node[pool] (lp) at (-1.72,0.05)
    {\textbf{Global file pool}\\28 files\\[4pt]
     Compare all files together};
  \draw[arr] (la.east) -- (lp.west);
  \draw[arr] (lb.east) -- (lp.west);
  \draw[arr] (lc.east) -- (lp.west);

  \node[scope] (ra) at (1.92,1.4)
    {\textbf{Module A}\\$\mathcal{F}_A$: 18 files\\select up to $K_s$};
  \node[scope] (rb) at (1.92,0)
    {\textbf{Module B}\\$\mathcal{F}_B$: 7 files\\select up to $K_s$};
  \node[scope] (rc) at (1.92,-1.4)
    {\textbf{Module C}\\$\mathcal{F}_C$: 3 files\\select up to $K_s$};
  \node[combined] (rp) at (5.08,0.05)
    {\textbf{File candidates}\\
     $\mathcal{W}_A,\mathcal{W}_B,\mathcal{W}_C$\\[3pt]
     $\mathcal{W}=\bigcup_m\mathcal{W}_m$\\[3pt]
     \textbf{$\downarrow$}\\File-Level\\Bug Isolation};
  \draw[arr] (ra.east) -- (rp.west);
  \draw[arr] (rb.east) -- (rp.west);
  \draw[arr] (rc.east) -- (rp.west);
\end{tikzpicture}
  \caption{The \emph{Scoped Candidate Competition} mechanism compared with naive global competition.}
  \Description{Two panels compare global and module-scoped file selection using three module candidates with 18, 7, and 3 files. The left panel pools all 28 files for comparison. The right panel compares files separately within each module and selects up to K-s file candidates per module. The three candidate sets are combined into W and passed to File-Level Bug Isolation without a global file ranking.}
  \label{fig:scoped-competition}
\end{figure}
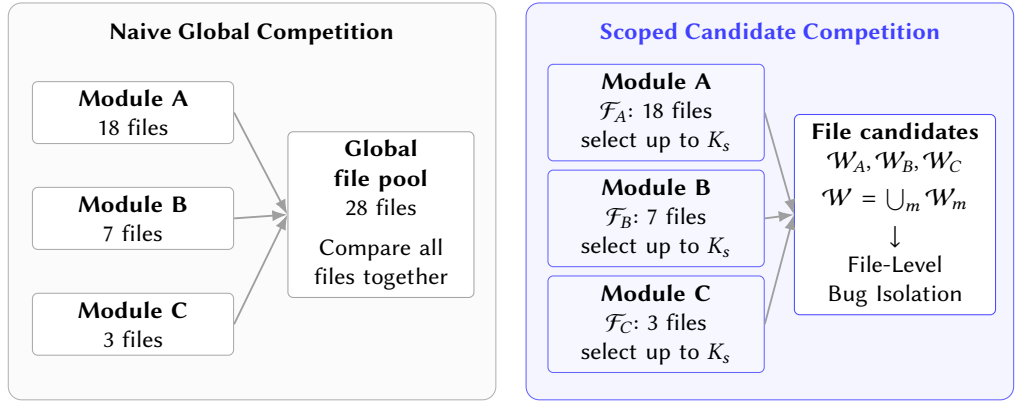

The \emph{Scoped Candidate Competition} mechanism limits the number of file candidates considered together by comparing files separately within each module candidate.  Comparing all files across module candidates in one prompt would increase the context size and could allow larger modules to dominate the candidate set.  For each module candidate $m\in\mathcal{S}_m$, the LLM analyzes its file set $\mathcal{F}_m$ using the file docs, module doc, bug context, and module-level assessment.  It compares the compiler functionality described in the file docs to assess whether each file may be responsible for the bug and ranks the files within that module.  It selects up to $K_s$ highest-ranked files as the candidate set
\begin{equation}
  \mathcal{W}_m=
  \operatorname{TopK}_{K_s}(\mathcal{F}_m),
\end{equation}
where $K_s$ is the maximum number of file candidates per module, and $\mathcal{W}_m$ contains the file candidates selected from module $m$.  The mechanism combines the file candidates from all module candidates into
\begin{equation}
  \mathcal{W}=
  \bigcup_{m\in\mathcal{S}_m}\mathcal{W}_m.
\end{equation}
The set $\mathcal{W}$ contains file candidates from different modules without a global file ranking and is passed to the \emph{File-Level Bug Isolation} component for further analysis.  As shown in Figure~\ref{fig:scoped-competition}, selecting files separately within each module limits the candidates considered together while preserving candidates from multiple modules.

\subsubsection{File-Level Bug Isolation}
\label{sec:file-isolation}

The third component, \emph{File-Level Bug Isolation}, produces the final ranked list of suspicious files.  It takes the file candidates $\mathcal{W}$ from the \emph{Module-Level Bug Isolation} component as its initial candidate set.  It first uses the \emph{Evidence-Guided Code Slicing} mechanism to extract code slices from each candidate file based on the bug context, limiting the amount of source code examined.  It then performs \emph{File-Level Bug Relevance Analysis}, using the code slices, file docs, bug context, crate-level assessments, and module-level assessments to assess whether each file may be responsible for the bug.  Finally, it performs \emph{Final File Ranking}, comparing these assessments to rank the files and return the most suspicious files.  The \emph{Evidence-Guided Code Slicing} mechanism consists of three steps, \emph{Evidence-Term Extraction}, \emph{Code Window Scoring}, and \emph{Top Window Selection}, which we describe below.

During \emph{Evidence-Term Extraction}, the mechanism extracts terms from the bug context $\mathcal{B}$, such as identifiers, diagnostic messages, and query names, to construct a weighted term set $\mathcal{T}$.  Terms from the compiler log and query trace receive weight $2$, whereas terms from the issue description and failure program receive weight $1$, giving greater weight to execution evidence.  For terms appearing in multiple sources, it uses the highest weight.  It also removes overly general terms that cannot distinguish source locations.  The resulting set $\mathcal{T}$ is used in \emph{Code Window Scoring}.

During \emph{Code Window Scoring}, the mechanism scores source lines in each candidate file using the weighted terms in $\mathcal{T}$.  For a source line $l$, the score is
\begin{equation}
  s(l) =
  \sum_{t \in \mathcal{T}}
  \mathbb{I}(t \in l)\,w(t) + b(l),
\end{equation}
where $\mathbb{I}(t\in l)$ indicates whether term $t$ occurs in line $l$, and $w(t)$ is the term's weight.  The bonus $b(l)$ increases the score of lines containing failure markers such as \texttt{panic!} or \texttt{assert!}.  Each line matching an evidence term is grouped with its neighboring lines to form a code window $W$.  The window score is the sum of its line scores:
\begin{equation}
  s(W)=\sum_{l\in W}s(l).
\end{equation}
Windows receive higher scores when they match more terms, match terms with greater weights, or contain more failure markers.  These scores guide \emph{Top Window Selection}.

During \emph{Top Window Selection}, the mechanism ranks each file's code windows by $s(W)$.  It selects the windows with the highest scores within a predefined budget for the amount of code extracted from each file.  The selected code is presented in source order with the file path and line-number ranges, forming the code slices used in \emph{File-Level Bug Relevance Analysis}.  Files without matching evidence terms are analyzed and ranked using their file docs and other available evidence, without code slices.  Figure~\ref{fig:code-slicing} illustrates this procedure for Bug 118990.

\begingroup
\setlength{\intextsep}{10pt}
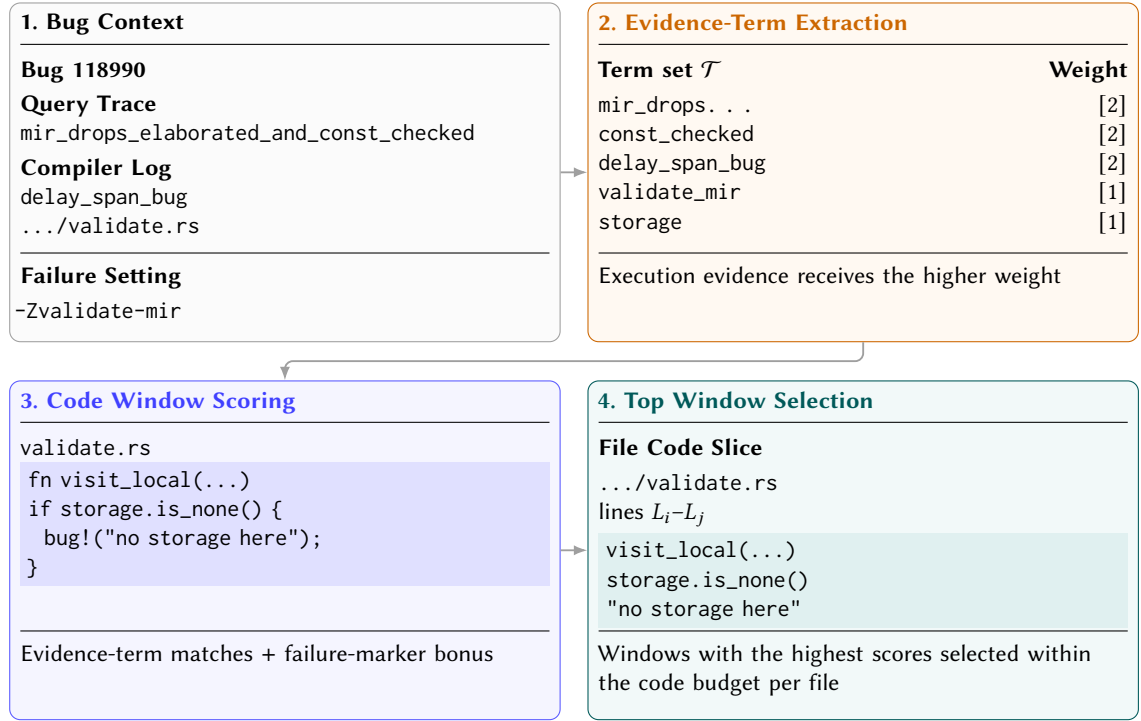
\begin{figure}[t]
  \centering
  \begin{tikzpicture}[
  font=\sffamily\normalsize,
  >={Latex[length=1.8mm]},
  arr/.style={->,draw=gray!75,line width=0.7pt},
  panel/.style={rounded corners=4pt,align=left,inner sep=4pt,
    anchor=north west}
]
  \node[panel,draw=gray!75,fill=gray!3,text width=7.0cm]
    (bug) at (0,0)
    {\begin{minipage}[t][4.2cm][t]{\linewidth}
     \raggedright
     \parbox[t][0.4cm][t]{\linewidth}{\textbf{1. Bug Context}}\par
     \hrule\vspace{6pt}
     \textbf{Bug 118990}\par\smallskip
     \textbf{Query Trace}\par
     \texttt{mir\_drops\_elaborated\_and\_const\_checked}\par\smallskip
     \textbf{Compiler Log}\par
     \texttt{delay\_span\_bug}\par
     \texttt{.../validate.rs}\par
     \vfill\hrule\vspace{5pt}
     \parbox[t][0.85cm][t]{\linewidth}{%
       \textbf{Failure Setting}\par\smallskip
       \texttt{-Zvalidate-mir}}
     \end{minipage}};

  \node[panel,draw=orange!80!black,fill=orange!5,text width=7.0cm]
    (terms) at (7.65,0)
    {\begin{minipage}[t][4.2cm][t]{\linewidth}
     \raggedright
     \parbox[t][0.4cm][t]{\linewidth}{\textbf{\color{orange!80!black} 2. Evidence-Term Extraction}}\par
     \hrule\vspace{6pt}
     \textbf{Term set $\mathcal{T}$}\hfill\textbf{Weight}\par\smallskip
     \texttt{mir\_drops\ldots}\hfill $[2]$\par
     \texttt{const\_checked}\hfill $[2]$\par
     \texttt{delay\_span\_bug}\hfill $[2]$\par
     \texttt{validate\_mir}\hfill $[1]$\par
     \texttt{storage}\hfill $[1]$\par
     \vfill\hrule\vspace{5pt}
     \parbox[t][0.85cm][t]{\linewidth}{%
       \raggedright Execution evidence receives the higher weight}
     \end{minipage}};

  \node[panel,draw=blue!65,fill=blue!4,text width=7.0cm]
    (score) at (0,-5.0)
    {\begin{minipage}[t][4.2cm][t]{\linewidth}
     \raggedright
     \parbox[t][0.4cm][t]{\linewidth}{\textbf{\color{blue!75} 3. Code Window Scoring}}\par
     \hrule\vspace{6pt}
     \texttt{validate.rs}\par\smallskip
     \colorbox{blue!12}{\parbox{\dimexpr\linewidth-2\fboxsep\relax}{%
       \ttfamily\raggedright
       fn visit\_local(...)\\
       if storage.is\_none() \{\\
       \ \ bug!("no storage here");\\
       \}}}\par
     \vfill\hrule\vspace{5pt}
     \parbox[t][0.85cm][t]{\linewidth}{%
       Evidence-term matches $+$ failure-marker bonus}
     \end{minipage}};

  \node[panel,draw=teal!70!black,fill=teal!5,text width=7.0cm]
    (selected) at (7.65,-5.0)
    {\begin{minipage}[t][4.2cm][t]{\linewidth}
     \raggedright
     \parbox[t][0.4cm][t]{\linewidth}{\textbf{\color{teal!70!black} 4. Top Window Selection}}\par
     \hrule\vspace{6pt}
     \textbf{File Code Slice}\par\smallskip
     \texttt{.../validate.rs}\par
     lines $L_i$--$L_j$\par\smallskip
     \colorbox{teal!12}{\parbox{\dimexpr\linewidth-2\fboxsep\relax}{%
       \ttfamily\raggedright
       visit\_local(...)\\
       storage.is\_none()\\
       "no storage here"}}\par
     \vfill\hrule\vspace{5pt}
     \parbox[t][0.85cm][t]{\linewidth}{%
       \raggedright Windows with the highest scores selected within the code budget per file}
     \end{minipage}};

  \draw[arr] (bug.east) -- (terms.west);
  \draw[arr,rounded corners=3pt] (terms.south) -- ++(0,-0.25) -| (score.north);
  \draw[arr] (score.east) -- (selected.west);
\end{tikzpicture}
  \caption{\emph{Evidence-Guided Code Slicing} illustrated with Bug 118990.}
  \Description{Four numbered panels in a two-by-two layout illustrate code slicing for Bug 118990. Arrows connect Bug Context, Evidence-Term Extraction, Code Window Scoring, and Top Window Selection in order. The bug context provides query-trace evidence, compiler-log evidence, and a compiler flag. Extracted terms receive weights of 1 or 2. A code window in validate.rs is scored using term matches and a bonus for failure markers. The selected code slice includes the file path, line-number range, and excerpts containing visit-local, storage.is-none, and the no storage here check.}
  \label{fig:code-slicing}
\end{figure}
\endgroup

During \emph{File-Level Bug Relevance Analysis}, the LLM analyzes each file $f\in\mathcal{W}$ using its file doc, available code slices, the bug context $\mathcal{B}$, crate-level assessments, and module-level assessments.  Based on the bug context and these assessments, it examines the compiler functionality described in the file doc and the implementation details in the available code slices to assess whether the file may be responsible for the bug.  For each file, the LLM produces an assessment for use in \emph{Final File Ranking}.

During \emph{Final File Ranking}, the LLM compares the file-level assessments to rank all files in $\mathcal{W}$ across modules.  Files considered more likely to be responsible for the bug are ranked higher.  It selects the top-$K_f$ files to form the final ranked list
\begin{equation}
  \mathcal{S}_f=
  \operatorname{TopK}_{K_f}(\mathcal{W}),
\end{equation}
where $K_f$ is the number of suspicious files returned, and $\operatorname{TopK}_{K_f}$ selects the $K_f$ highest-ranked files based on these assessments.  HIDE returns the ranked list $\mathcal{S}_f$ with a brief reason for selecting each file.  This list corresponds to the suspicious files in Figure~\ref{fig:approach-overview}.

\begin{figure}[t]
  \centering
  \begin{tikzpicture}[x=1cm,y=1cm]
  \definecolor{promptborder}{HTML}{5B6F8A}
  \definecolor{promptfill}{HTML}{F7F9FC}
  \definecolor{prompthead}{HTML}{E8EEF7}
  \definecolor{promptlabel}{HTML}{315E9B}

  \node[anchor=north west,draw=promptborder,rounded corners=4pt,
        line width=0.8pt,fill=promptfill,inner sep=0pt,text=black]
    at (0,0) {%
    \begin{minipage}{14.8cm}
      \colorbox{prompthead}{%
        \parbox[c][0.6cm][c]{\dimexpr\linewidth-2\fboxsep\relax}{%
          \centering\sffamily\bfseries\normalsize\color{promptborder}
          Prompt Template}}\par
      \vspace{6pt}
      \noindent\hspace*{0.4cm}%
      \begin{minipage}{14.0cm}
      \ttfamily\normalsize\raggedright
      You are isolating the source files responsible for a rustc compiler
      bug.
      \\[0.55em]
      \textcolor{promptlabel}{Isolation level:}
      \textless level\textgreater\\
      \textcolor{promptlabel}{Bug context:}
      \textless bug\_context\textgreater\\
      \textcolor{promptlabel}{Upstream results:}
      \textless upstream\_results\textgreater\\
      \textcolor{promptlabel}{Candidate documentation:}
      \textless candidate\_docs\textgreater\\
      \textcolor{promptlabel}{Code evidence:}
      \textless code\_slices\textgreater\\
      \textcolor{promptlabel}{Output size:}
      \textless K\textgreater
      \\[0.55em]
      \textcolor{promptlabel}{Instructions:}\\
      Analyze how each candidate's responsibility relates to the observed
      failure. Distinguish candidates that directly expose the symptom,
      intermediate compiler phases that propagate it, and upstream
      components that may introduce the root cause.
      \\[0.55em]
      \textcolor{promptlabel}{Output:}\\
      Return exactly \textless K\textgreater{} valid candidate IDs, or up to
      \textless K\textgreater{} for Scoped Candidate Competition. Order them
      from most to least relevant to the bug's root cause. For each
      candidate, provide a concise
      reason grounded in the bug context and the supplied documentation or
      code evidence.
      \end{minipage}\par
      \vspace{8pt}
    \end{minipage}};
\end{tikzpicture}
  \caption{Prompt template for \emph{Hierarchical Bug Isolation}.}
  \Description{A framed prompt template presents a task statement, input fields, instructions, and an output requirement. The fields specify the isolation level, bug context, upstream results, candidate documentation, code slices, and output size K. The instructions ask the LLM to distinguish candidates that expose the symptom, propagate it through intermediate compiler phases, or introduce the root cause. The output is exactly K valid candidate identifiers, or up to K for Scoped Candidate Competition, ranked by relevance to the root cause with a brief reason for each candidate based on the supplied evidence.}
  \label{fig:isolation-prompt}
  \par\medskip
  \begin{minipage}{\linewidth}
  \captionof{table}{Prompt design across the three isolation components and the scoped candidate competition mechanism.}
  \label{tab:isolation-prompts}
  \centering
  \small
  \setlength{\tabcolsep}{2pt}
  \renewcommand{\arraystretch}{1.15}
  \begin{tabular}{@{}>{\raggedright\arraybackslash}p{0.24\linewidth}
                      >{\raggedright\arraybackslash}p{0.29\linewidth}
                      >{\raggedright\arraybackslash}p{0.275\linewidth}
                      >{\raggedright\arraybackslash}p{0.155\linewidth}@{}}
    \toprule
    Component or mechanism & Input content & Processing & Output \\
    \midrule
    \emph{Crate-Level Bug Isolation}
      & Bug context and crate docs
      & \emph{Crate-Level Bug Relevance Analysis}\par
        \emph{Crate Candidate Ranking}
      & $K_c$ crates \\
    \midrule
    \emph{Module-Level Bug Isolation}
      & Bug context, crate-level\newline assessments, and module docs
      & \emph{Module-Level Bug Relevance Analysis}\par
        \emph{Module Candidate Ranking}
      & $K_m$ modules \\
    \midrule
    \emph{Scoped Candidate Competition}
      & Bug context, module assessment,\newline module doc, and file docs
      & \emph{Scoped Candidate Competition}
      & Up to $K_s$ files \\
    \midrule
    \emph{File-Level Bug Isolation}
      & Bug context, assessments,\newline file docs, and code slices
      & \emph{File-Level Bug Relevance Analysis}\par
        \emph{Final File Ranking}
      & $K_f$ suspicious files \\
    \bottomrule
  \end{tabular}
  \end{minipage}
\end{figure}

\paragraph{Prompt Design.}
The three components and the \emph{Scoped Candidate Competition} mechanism use the shared prompt template in Figure~\ref{fig:isolation-prompt}, with their inputs, processing, and outputs summarized in Table~\ref{tab:isolation-prompts}.  The template specifies the isolation level, bug context, candidate docs, available assessments and code slices, and output size.  At the crate level, the input contains the bug context and crate docs, and the prompt asks whether the compiler functionality described in each doc may be responsible for the bug.  At the module level, the input contains module docs and crate-level assessments alongside the bug context, and the prompt asks the same question for modules within the selected crates.  For the \emph{Scoped Candidate Competition} mechanism, the input contains the bug context, one module candidate's assessment and doc, and its file docs, and the prompt asks for file comparison within that module.  At the file level, the input contains the bug context, crate-level assessments, module-level assessments, file docs, and available code slices from the \emph{Evidence-Guided Code Slicing} mechanism, and the prompt asks for analysis of the combined file candidates using both compiler functionality and implementation details.  For each component or mechanism, the LLM follows the processing listed in Table~\ref{tab:isolation-prompts} to rank candidates and return the specified number, or up to $K_s$ for \emph{Scoped Candidate Competition}, with a brief reason for selecting each candidate.

In summary, the \emph{Hierarchical Bug Isolation} stage uses the generated docs and bug context to narrow candidates across crates, modules, and files, producing a ranked list of suspicious files.  The \emph{Scoped Candidate Competition} and \emph{Evidence-Guided Code Slicing} mechanisms limit the number of candidates and the amount of source code included in each prompt, respectively.

\section{Experimental Setup}
\label{sec:experimental-setup}

To evaluate the performance of HIDE in compiler bug isolation, we investigate the following research questions:
\begin{itemize}
  \item \textbf{RQ1 (Effectiveness):} How effective is HIDE in isolating rustc bugs compared with baselines?

  \item \textbf{RQ2 (Efficiency):} How efficient is HIDE compared with baselines?

  \item \textbf{RQ3 (Generalizability):} How does HIDE perform across different LLMs?

  \item \textbf{RQ4 (Ablation):} How do the components and mechanisms of HIDE contribute to its performance?
\end{itemize}

\subsection{Datasets}
\label{sec:datasets}

Our evaluation uses the rustc bug dataset assembled by Liu et al.~\cite{liu2025empirical}.  Since the dataset was constructed for an empirical study of rustc bugs, it requires further processing for file-level bug isolation.  We process the dataset in four steps: defining the target bug categories, screening bugs for artifact availability and reproducibility, generating the artifacts required by the evaluated methods, and deriving the file-level ground truth.  We describe each step below.

First, we focus our evaluation on ICE, completeness, and soundness bugs.  An ICE bug causes rustc to terminate with an internal compiler error.  A completeness bug causes rustc to reject a valid program, while a soundness bug causes it to accept an invalid program.  These three categories admit a uniform compiler-level reproduction oracle based solely on compiler termination and program acceptance.  In contrast, runtime-behavior bugs require executing the compiled program under a controlled runtime environment, while diagnostic bugs require specialized oracles for diagnostic messages, source spans, or suggested fixes.  We therefore exclude runtime-behavior and diagnostic bugs from our evaluation to avoid confounding bug-isolation effectiveness with differences in reproduction artifacts and oracle design.  This exclusion serves only to delimit the scope of our evaluation and does not imply an inherent limitation on the applicability of HIDE.

Second, we screen all 195 bugs in these three categories from Liu et al.~\cite{liu2025empirical} for artifact availability and reproducibility.  Each bug must meet three inclusion criteria.  (1) The issue description provides enough information to construct a failure program.  (2) The rustc revision and compilation command, including the required parameters, can be identified.  (3) Compiling the failure program on that revision consistently reproduces the reported behavior.  This screening excludes 35 bugs with missing information or inconsistent behavior, leaving 160 bugs in the evaluation dataset.

Third, we prepare the artifacts required for bug isolation.  For each bug in the evaluation dataset, we collect the issue description, failure program, compilation parameters, and rustc revision on which the bug is reproduced.  We compile the failure program on this revision with the specified parameters to collect the compiler log.  During the same compilation, rustc's self-profile mechanism records the executed queries, which we use to construct the query trace.  The issue description, failure program, compiler log, and query trace form the bug context.  We use the fixing pull request (PR) only to derive the ground truth and do not use its information to generate the bug context.

Finally, we derive the file-level ground truth for each bug from its fixing PR.  We include the Rust source files in the rustc compiler source tree modified by the PR and exclude test files, documentation, and other non-compiler source files.  A bug may have multiple ground-truth files if its fixing PR modifies several compiler source files.  We consider a bug successfully isolated when at least one ground-truth file appears in the returned top-$K$ suspicious files.

\begin{table}[t]
  \caption{Characteristics of the rustc bugs used in our evaluation.}
  \label{tab:dataset-distribution}
  \centering
  \small
  \setlength{\tabcolsep}{5pt}
  \begin{tabular}{@{}lrcc@{}}
    \toprule
    Category & Bugs & rustc \texttt{.rs} files & rustc versions \\
    \midrule
    ICE          & 94  & 1,258--1,863 & 1.53.0--1.85.0 \\
    Completeness & 46  & 1,377--1,858 & 1.60.0--1.84.0 \\
    Soundness    & 20  & 1,406--1,787 & 1.61.0--1.80.0 \\
    \midrule
    Total        & 160 & 1,258--1,863 & 1.53.0--1.85.0 \\
    \bottomrule
  \end{tabular}
\end{table}

Table~\ref{tab:dataset-distribution} reports the number of bugs and the ranges of rustc source-file counts and compiler versions in each category.  The file counts correspond to the rustc revision associated with each bug.

\subsection{Baselines}
\label{sec:baselines}

We compare HIDE with three state-of-the-art baselines for general software bug isolation, namely Agentless~\cite{xia2025demystifying}, SoapFL~\cite{qin2025s}, and GraphLocator~\cite{liu2026graphlocator}.  These methods represent \emph{semantics-based}, \emph{execution-based}, and \emph{dependency-based} approaches, respectively.  We evaluate their ability to rank suspicious rustc source files.

\begin{itemize}
  \item \textbf{Agentless} is the \emph{semantics-based} baseline.  Its hierarchical localization process progressively narrows an issue from suspicious files to relevant classes or functions, followed by specific edit locations.  At the file level, it combines an LLM's analysis of the repository tree with embedding retrieval over chunks of source code.  Subsequent levels use code skeletons and selected source code for more detailed localization.  We evaluate its localization results without performing the repair and validation stages.

  \item \textbf{SoapFL} is the \emph{execution-based} baseline.  It organizes fault localization into fault comprehension, codebase navigation, and fault confirmation.  During fault comprehension, it analyzes the behavior of failing tests to infer possible causes.  During codebase navigation, it combines coverage with existing documentation and documentation generated by an LLM to narrow covered classes and methods.  During fault confirmation, it reviews candidate methods through multiple LLM rounds to produce a suspiciousness ranking.

  \item \textbf{GraphLocator} is the \emph{dependency-based} baseline.  It represents repository entities and their hierarchical and dependency relations using a repository dependency fractal structure.  It first locates graph vertices associated with reported symptoms, then traverses neighboring vertices and uses abductive reasoning to construct a causal issue graph that grounds sub-issues in code entities and records their causal dependencies.  This graph supports tracing reported symptoms toward potential bug causes.
\end{itemize}

To ensure a consistent comparison, we preserve each baseline's core localization workflow, prompt design, and reported parameter settings.  We evaluate all methods using a shared dataset and protocol, comparing their ranked lists of suspicious files.

\subsection{Evaluation Metrics}
\label{sec:evaluation-metrics}

We evaluate file-level bug isolation effectiveness using Accuracy@$k$ and computational efficiency using average time and average tokens.

Following prior bug-isolation studies~\cite{li2025llm,qin2025s,tu2024llm4cbi}, we define Accuracy@$k$ as the proportion of bugs for which at least one ground-truth file appears among the top-$k$ suspicious files returned by a method.  For bug $i$, let $P_i^k$ be the set of these top-$k$ files and $G_i$ be the ground-truth file set.  Accuracy@$k$ is calculated as
\begin{equation}
  \operatorname{Accuracy@}k =
  \frac{1}{N}
  \sum_{i=1}^{N}
  \mathbb{I}\left(P_i^k \cap G_i \neq \emptyset\right),
  \label{eq:accuracy-at-k}
\end{equation}
where $N$ is the number of evaluated bugs and $\mathbb{I}(\cdot)$ equals one if the condition holds and zero otherwise.  We report results for $k \in \{1,3,5,10\}$, with higher values of Accuracy@$k$ indicating better bug isolation effectiveness.

Average time is the mean wall-clock time per bug, measured from the start of bug isolation until the final ranked list is produced.  Average tokens is the mean total number of input and output tokens across all LLM requests per bug.  Lower values indicate less time and fewer tokens required for bug isolation, respectively.

\subsection{Experimental Design}
\label{sec:experimental-design}

For RQ1 and RQ2, HIDE and all baselines use GPT-5.4.  For RQ4, the full approach and all ablated variants also use GPT-5.4.  These settings keep the underlying LLM consistent when comparing methods and ablated variants.  For RQ3, we evaluate HIDE with GPT-5.4, Gemini-3.1-Pro-Preview, and DeepSeek-V4-Pro to examine its performance across LLMs.  We perform three independent runs for each method in RQ1 and each model configuration in RQ3, and report the mean Accuracy@$k$ values.

\subsection{Implementation Details}
\label{sec:implementation-details}

Table~\ref{tab:implementation-parameters} lists the parameters that control candidate selection in the \emph{Hierarchical Bug Isolation} stage.  We select their values using GPT-5.4 on a subset comprising 20\% of the full dataset.

We evaluate every combination of the candidate values for $K_c$, $K_m$, and $K_s$ in Table~\ref{tab:implementation-parameters}.  We select the configuration with the highest Accuracy@10, using lower average tokens and then lower average time to break ties.  We fix $K_f$ at 10 to support all reported Accuracy@$k$ metrics.

\begin{table}[t]
  \caption{Candidate values and selected parameter settings.}
  \label{tab:implementation-parameters}
  \centering
  \small
  \setlength{\tabcolsep}{5pt}
  \begin{tabular}{@{}clcc@{}}
    \toprule
    Parameter & Meaning & Candidate values & Selected value \\
    \midrule
    $K_c$ & Number of selected crate candidates & 2, 4, 6 & 4 \\
    $K_m$ & Number of selected module candidates & 8, 12, 16 & 12 \\
    $K_s$ & Maximum number of files selected per module & 2, 4, 6 & 4 \\
    $K_f$ & Number of returned suspicious files & Fixed & 10 \\
    \bottomrule
  \end{tabular}
\end{table}

The selected configuration is $K_c=4$, $K_m=12$, $K_s=4$, and $K_f=10$.  We use these values across LLMs, the main experiments, and the ablation study.

All experiments run on the same Ubuntu 22.04.5 LTS server, equipped with Intel Xeon Platinum 8468V processors and 96 physical CPU cores.    

\section{Experimental Results and Analysis}
\label{sec:experimental-results}

\subsection{RQ1: Effectiveness}
\label{sec:rq1-effectiveness}

\paragraph{Experimental Results.}
Table~\ref{tab:effectiveness} compares HIDE with the baselines using mean Accuracy@$k$ values from three independent runs with GPT-5.4.  Boldface marks the best result, and underlining marks the second-best in each column.

\begin{table}[t]
  \caption{File-level bug isolation effectiveness with GPT-5.4.}
  \label{tab:effectiveness}
  \centering
  \small
  \setlength{\tabcolsep}{5pt}
  \begin{tabular}{@{}lcccc@{}}
    \toprule
    Approach & Accuracy@1 & Accuracy@3 & Accuracy@5 & Accuracy@10 \\
    \midrule
    HIDE & \textbf{42.08\%} & \textbf{60.21\%} &
      \textbf{70.83\%} & \textbf{80.42\%} \\
    Agentless & 30.00\% & \underline{55.21\%} &
      \underline{64.38\%} & \underline{74.38\%} \\
    SoapFL & 33.75\% & 47.50\% & 56.46\% & 67.50\% \\
    GraphLocator & \underline{36.25\%} & 48.75\% & 50.00\% & 50.00\% \\
    \bottomrule
  \end{tabular}
\end{table}

\underline{For overall accuracy}, HIDE achieves the highest accuracy at all four values of $k$.  Its Accuracy@1 is 42.08\%, exceeding the strongest baseline, GraphLocator, by 5.83\%.  Agentless is the strongest baseline at $k=3$, $5$, and $10$.  HIDE achieves 60.21\%, 70.83\%, and 80.42\% at these values of $k$, exceeding Agentless by 5.00\%, 6.45\%, and 6.04\%, respectively.

\underline{For accuracy gains}, HIDE's accuracy rises from 42.08\% at $k=1$ to 80.42\% at $k=10$, a gain of 38.34\%.  Increasing $k$ from 5 to 10 adds 9.59\%, while GraphLocator remains at 50.00\%.  These results indicate that inspecting more candidates in HIDE's ranked list improves the likelihood of identifying a ground-truth file.

\paragraph{Experimental Analysis.}

\underline{Comparison with Agentless.} Agentless has the lowest Accuracy@1 at 30.00\% but ranks second at Accuracy@3, Accuracy@5, and Accuracy@10.  Its file localization combines analysis of the repository tree with embedding retrieval over code chunks~\cite{xia2025demystifying}.  This retrieval emphasizes semantic relevance to the issue description rather than compiler execution behavior.  Agentless's accuracy rises from 30.00\% at Accuracy@1 to 55.21\% at Accuracy@3, suggesting that relevant files often appear in its ranked list but are not prioritized first.  HIDE instead compares documented file responsibilities with code slices selected using bug evidence during \emph{File-Level Bug Isolation}.  It achieves 42.08\% Accuracy@1, compared with 30.00\% for Agentless, an improvement of 12.08\%.  This integration helps HIDE prioritize the top candidate more effectively while retaining higher accuracy at larger values of $k$.

\underline{Comparison with SoapFL.} SoapFL ranks third among the four approaches at Accuracy@1, Accuracy@5, and Accuracy@10, whereas it ranks fourth at Accuracy@3.  Its \emph{Test Behavior Analysis} task summarizes test execution for the \emph{Test Failure Analysis} task to infer bug causes.  The \emph{Search Suspicious Class} task narrows covered classes, while the \emph{Method Doc Enhancement} task supports the \emph{Find Related Methods} task.  Repeated executions of the \emph{Method Review} task precede the \emph{Suspicious Method Ranking} task~\cite{qin2025s}.  This workflow uses test behavior and documentation to narrow candidates from classes to methods, but it lacks HIDE's aggregated descriptions of compiler functionality at the crate and module levels.  HIDE combines these descriptions with query traces to narrow compiler regions before file comparison.  It achieves 80.42\% Accuracy@10 compared with SoapFL's 67.50\%, an improvement of 12.92\%.  This result suggests that compiler level context helps preserve relevant files in the final ranking.

\underline{Comparison with GraphLocator.} GraphLocator ranks second at Accuracy@1 (36.25\%), but falls to third at Accuracy@3 and fourth at Accuracy@5 and Accuracy@10.  Its \emph{Dynamic CIG Discovering} phase traces from symptom locations toward potential causes through neighboring code entities~\cite{liu2026graphlocator}.  This causal reasoning provides a basis for prioritizing bug causes rather than relying on semantic similarity alone.  However, its Accuracy@5 and Accuracy@10 both remain at 50.00\%, so extending the ranked list beyond the top five candidates does not improve its localization accuracy.  This pattern is consistent with search guided by code dependencies concentrating successful predictions near the top without extending coverage of ground truth files at larger $k$.  HIDE's \emph{Scoped Candidate Competition} retains alternatives from separate modules before global ranking, with Accuracy@5 to Accuracy@10 rising from 70.83\% to 80.42\%.

\rqanswer{RQ1}{%
HIDE outperforms all baselines at every reported Accuracy@$k$, achieving 42.08\% Accuracy@1 and 80.42\% Accuracy@10.  Although HIDE consistently exceeds the baselines, its Accuracy@1 leaves substantial room for improvement.  Accurately ranking a ground-truth file first therefore remains a challenge.}

\subsection{RQ2: Efficiency}
\label{sec:rq2-efficiency}

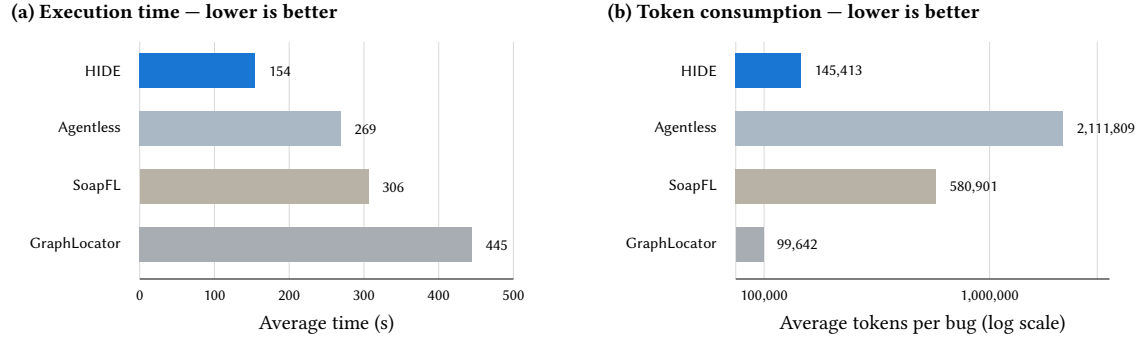
\begin{figure}[t]
  \centering
  \resizebox{\columnwidth}{!}{\begin{tikzpicture}[x=1cm,y=1cm,font=\sffamily\footnotesize]
  \definecolor{effmain}{HTML}{1976D2}
  \definecolor{effbasea}{HTML}{AAB7C4}
  \definecolor{effbaseb}{HTML}{B8B1A5}
  \definecolor{effbasec}{HTML}{A7ADB3}
  \definecolor{effgrid}{HTML}{D9D9D9}

  \node[anchor=west,font=\bfseries\normalsize] at (0,4.55)
    {(a) Execution time --- lower is better};
  \foreach \x/\lab in {2.10/0,3.26/100,4.42/200,5.58/300,6.74/400,7.90/500} {
    \draw[effgrid] (\x,0.42) -- (\x,4.10);
    \node[anchor=north] at (\x,0.35) {\lab};
  }
  \node[anchor=east] at (1.92,3.65) {HIDE};
  \node[anchor=east] at (1.92,2.75) {Agentless};
  \node[anchor=east] at (1.92,1.85) {SoapFL};
  \node[anchor=east] at (1.92,0.95) {GraphLocator};
  \fill[effmain]  (2.10,3.38) rectangle (3.89,3.92);
  \fill[effbasea] (2.10,2.48) rectangle (5.22,3.02);
  \fill[effbaseb] (2.10,1.58) rectangle (5.65,2.12);
  \fill[effbasec] (2.10,0.68) rectangle (7.26,1.22);
  \node[anchor=west] at (3.99,3.65) {154};
  \node[anchor=west] at (5.32,2.75) {269};
  \node[anchor=west] at (5.75,1.85) {306};
  \node[anchor=west] at (7.36,0.95) {445};
  \draw (2.10,0.42) -- (7.90,0.42);
  \node[anchor=north,font=\normalsize] at (5.00,-0.02) {Average time (s)};

  \node[anchor=west,font=\bfseries\normalsize] at (9.25,4.55)
    {(b) Token consumption --- lower is better};
  \foreach \x in {11.35,11.79,15.29,16.96} {
    \draw[effgrid] (\x,0.42) -- (\x,4.10);
  }
  \node[anchor=north] at (11.79,0.35) {100,000};
  \node[anchor=north] at (15.29,0.35) {1,000,000};
  \node[anchor=east] at (11.17,3.65) {HIDE};
  \node[anchor=east] at (11.17,2.75) {Agentless};
  \node[anchor=east] at (11.17,1.85) {SoapFL};
  \node[anchor=east] at (11.17,0.95) {GraphLocator};
  \fill[effmain]  (11.35,3.38) rectangle (12.36,3.92);
  \fill[effbasea] (11.35,2.48) rectangle (16.43,3.02);
  \fill[effbaseb] (11.35,1.58) rectangle (14.46,2.12);
  \fill[effbasec] (11.35,0.68) rectangle (11.78,1.22);
  \node[anchor=west] at (12.46,3.65) {145,413};
  \node[anchor=west] at (16.53,2.75) {2,111,809};
  \node[anchor=west] at (14.56,1.85) {580,901};
  \node[anchor=west] at (11.88,0.95) {99,642};
  \draw (11.35,0.42) -- (17.15,0.42);
  \node[anchor=north,font=\normalsize] at (14.25,-0.02)
    {Average tokens per bug (log scale)};
\end{tikzpicture}}
  \caption{Efficiency of HIDE and the baselines. (a) Average time per bug (seconds). (b) Average tokens per bug (logarithmic scale).}
  \Description{Two horizontal bar charts compare average time per bug in seconds and average tokens per bug on a logarithmic scale. HIDE has the lowest average time. GraphLocator uses the fewest tokens, while Agentless uses the most.}
  \label{fig:efficiency}
\end{figure}

\paragraph{Experimental Results.}
Figure~\ref{fig:efficiency} compares HIDE with the baselines in terms of average time and average tokens per bug.

\underline{For average time}, HIDE is the fastest approach, requiring 154 seconds per bug.  Agentless, SoapFL, and GraphLocator require 269, 306, and 445 seconds per bug, respectively.  HIDE reduces average time by approximately 42.8\%, 49.7\%, and 65.4\% relative to these baselines, respectively.

\underline{For average tokens}, HIDE uses 145,413 tokens per bug, compared with 2,111,809 for Agentless and 580,901 for SoapFL, corresponding to reductions of approximately 93.1\% and 75.0\%, respectively.  GraphLocator uses the fewest tokens at 99,642 per bug.  Compared with GraphLocator, HIDE uses 45,771 more tokens but requires 291 fewer seconds per bug.

\paragraph{Experimental Analysis.}

\underline{Comparison with Agentless.} Agentless processes 2,111,809 tokens per bug, about 14.52 times HIDE's 145,413 tokens.  Its \emph{Localize to Related Elements} step places the skeletons of all candidate files in one prompt, while the \emph{Localize to Edit Locations} step repeatedly supplies selected source code when sampling edit locations~\cite{xia2025demystifying}.  These repeated inputs create cumulative token costs despite skeleton compression.  HIDE narrows candidates using layered documentation before detailed file analysis.  Its \emph{Scoped Candidate Competition} limits the files compared together, while \emph{Evidence-Guided Code Slicing} restricts the source content supplied for analysis.  Together, these mechanisms limit the context HIDE processes, explaining its substantially lower token consumption than Agentless.

\underline{Comparison with SoapFL.} SoapFL requires 306 seconds per bug, nearly twice HIDE's 154 seconds.  Its 580,901 tokens are about 3.99 times HIDE's consumption.  SoapFL's \emph{Method Doc Enhancement} task reads method code to generate descriptions, while its \emph{Method Review} task repeatedly resubmits failure context for each candidate method~\cite{qin2025s}.  These tasks add LLM requests involving both source code and test evidence, increasing cumulative processing.  HIDE instead compares selected files using bounded code slices, without a separate review dialogue for each method.

\underline{Comparison with GraphLocator.} GraphLocator uses the fewest tokens at 99,642 per bug, but requires 445 seconds, about 2.89 times HIDE's execution time.  GraphLocator's \emph{Symptom Vertices Locating} phase iteratively retrieves code entities.  Its \emph{Dynamic CIG Discovering} phase then expands one subissue per iteration using newly observed code entities alongside a compact causal graph~\cite{liu2026graphlocator}.  Although this process restricts source context, each expansion depends on the previous reasoning result.  These operations explain why low token consumption does not translate into a short execution time.  HIDE follows a fixed hierarchy, using 45,771 more tokens but finishing 291 seconds sooner.

\rqanswer{RQ2}{%
HIDE achieves the lowest average time at 154 seconds per bug and uses 145,413 tokens per bug.  Its average token consumption is substantially lower than that of Agentless and SoapFL but higher than that of GraphLocator.}

\subsection{RQ3: Generalizability}
\label{sec:rq3-generalizability}

\paragraph{Experimental Results.}
Table~\ref{tab:generalizability} reports HIDE's bug isolation accuracy over three independent runs with each of three LLMs: GPT-5.4, Gemini-3.1-Pro-Preview, and DeepSeek-V4-Pro.  We examine accuracy across models using the mean results and consistency across runs using the difference between the highest and lowest accuracy values for each model.  The settings for candidate selection remain unchanged across models.  Bold and underlined values in the Mean column indicate the best and second-best results for each metric, respectively.

\underline{For accuracy across models}, HIDE achieves similar mean Accuracy@10 values, reaching 80.42\% with GPT-5.4, 80.00\% with Gemini-3.1-Pro-Preview, and 76.04\% with DeepSeek-V4-Pro.  Differences are larger at Accuracy@1, Accuracy@3, and Accuracy@5, where Gemini-3.1-Pro-Preview achieves the highest means of 53.33\%, 69.17\%, and 75.21\%, respectively.  The largest difference between models is 14.58\% at Accuracy@1, compared with 4.38\% at Accuracy@10.

\underline{For consistency across runs}, HIDE exhibits the smallest variation with GPT-5.4 at every reported metric.  Across the three runs, the differences between its highest and lowest accuracy values are 1.25\% at Accuracy@1 and 2.50\% at Accuracy@10.  By comparison, the difference reaches 7.50\% at Accuracy@1 with Gemini-3.1-Pro-Preview and 6.88\% at Accuracy@10 with DeepSeek-V4-Pro.

\begin{table}[t]
  \caption{File-level bug isolation accuracy of HIDE across three LLMs and three independent runs.}
  \label{tab:generalizability}
  \centering
  \small
  \setlength{\tabcolsep}{4pt}
  \begin{tabular}{@{}llrrrr@{}}
    \toprule
    LLM & Metric & Run 1 & Run 2 & Run 3 & Mean \\
    \midrule
    \multirow{4}{*}{GPT-5.4}
      & Accuracy@1  & 42.50\% & 41.25\% & 42.50\% & \underline{42.08\%} \\
      & Accuracy@3  & 60.00\% & 61.25\% & 59.38\% & \underline{60.21\%} \\
      & Accuracy@5  & 70.00\% & 71.88\% & 70.62\% & \underline{70.83\%} \\
      & Accuracy@10 & 81.88\% & 80.00\% & 79.38\% & \textbf{80.42\%} \\
    \midrule
    \multirow{4}{*}{Gemini-3.1-Pro-Preview}
      & Accuracy@1  & 50.00\% & 52.50\% & 57.50\% & \textbf{53.33\%} \\
      & Accuracy@3  & 66.88\% & 68.12\% & 72.50\% & \textbf{69.17\%} \\
      & Accuracy@5  & 73.12\% & 75.00\% & 77.50\% & \textbf{75.21\%} \\
      & Accuracy@10 & 78.12\% & 80.00\% & 81.88\% & \underline{80.00\%} \\
    \midrule
    \multirow{4}{*}{DeepSeek-V4-Pro}
      & Accuracy@1  & 37.50\% & 40.00\% & 38.75\% & 38.75\% \\
      & Accuracy@3  & 53.75\% & 59.38\% & 55.62\% & 56.25\% \\
      & Accuracy@5  & 61.88\% & 65.62\% & 63.75\% & 63.75\% \\
      & Accuracy@10 & 73.12\% & 80.00\% & 75.00\% & 76.04\% \\
    \bottomrule
  \end{tabular}
\end{table}

\paragraph{Experimental Analysis.}

\underline{Regarding accuracy across models}, Gemini-3.1-Pro-Preview ranks first at Accuracy@1, Accuracy@3, and Accuracy@5, while GPT-5.4 ranks first at Accuracy@10 (80.42\%, narrowly ahead of Gemini at 80.00\%).  Gemini's 53.33\% Accuracy@1 indicates stronger prioritization of a responsible file at the top, potentially reflecting its software engineering reasoning~\cite{google2026gemini31pro}.  By contrast, GPT-5.4 gains 9.59\% from Accuracy@5 to Accuracy@10, compared with 4.79\% for Gemini, suggesting that it recovers more bugs as the candidate list expands.  This pattern may reflect GPT-5.4's coding and document understanding capabilities, which are useful for interpreting evidence across HIDE's documentation hierarchy~\cite{openai2026gpt54}.

\underline{Regarding consistency across runs}, GPT-5.4 has the smallest Accuracy@1 span across the three runs, ranging from 41.25\% to 42.50\%, a span of 1.25\%.  Gemini-3.1-Pro-Preview has the largest, ranging from 50.00\% to 57.50\%, a span of 7.50\%.  Because LLMs can produce different outputs from identical prompts~\cite{openai2026evaluation}, early differences in crate or module selection may alter the downstream candidate pool and final file ranking.  GPT-5.4 therefore shows more consistent Accuracy@1 than Gemini in these runs.  With only three runs, these spans describe observed variation under the tested settings and do not establish inherent differences in model randomness.

\rqanswer{RQ3}{%
HIDE maintains similar accuracy across models at $k=10$ under unchanged settings for candidate selection, with mean Accuracy@10 ranging from 76.04\% to 80.42\%, although differences are larger at smaller values of $k$.  Consistency across runs depends on the model, with GPT-5.4 showing the smallest accuracy variation across the three runs.}

\subsection{RQ4: Ablation Study}
\label{sec:rq4-ablation}

\paragraph{Experimental Results.}
Table~\ref{tab:ablation} compares HIDE with eight ablated variants, grouped into isolation components, documentation generation components, and context management mechanisms.  Bold values indicate the best result for each metric.  All eight variants achieve lower Accuracy@$k$ than the full approach at every reported value of $k$.

\underline{For isolation components}, removing the \emph{Crate-Level Bug Isolation} or \emph{Module-Level Bug Isolation} component reduces Accuracy@10 by 3.76\% or 8.13\%, respectively.  Removing both produces the largest Accuracy@10 decrease in this group, at 10.00\%, and also reduces Accuracy@1 by 5.62\%.

\underline{For documentation generation components}, removing the \emph{Crate-Level Documentation Generation}, \emph{Module-Level Documentation Generation}, or \emph{File-Level Documentation Generation} component reduces Accuracy@10 by 4.38\%, 5.63\%, or 13.13\%, respectively.  Removing the file-level component produces the largest Accuracy@10 decrease among all eight variants and the largest decreases at every reported metric within this group.

\underline{For context management mechanisms}, removing the \emph{Scoped Candidate Competition} mechanism reduces Accuracy@5 and Accuracy@10 by 9.38\% and 10.00\%, respectively.  Removing the \emph{Evidence-Guided Code Slicing} mechanism has its largest effect at Accuracy@10, with a decrease of 9.38\%, compared with decreases of 3.75\%, 0.62\%, and 1.88\% at Accuracy@1, Accuracy@3, and Accuracy@5, respectively.

\begin{table}[t]
  \caption{File-level bug isolation accuracy of HIDE and its ablated variants.}
  \label{tab:ablation}
  \centering
  \small
  \setlength{\tabcolsep}{3pt}
  \begin{tabular}{@{}>{\raggedright\arraybackslash}p{0.43\linewidth}cccc@{}}
    \toprule
    Variant & Accuracy@1 & Accuracy@3 & Accuracy@5 & Accuracy@10 \\
    \midrule
    \textbf{HIDE (Full)} & \textbf{42.50\%} & \textbf{60.00\%} &
      \textbf{70.00\%} & \textbf{81.88\%} \\
    \midrule
    \multicolumn{5}{l}{\textit{Isolation components}} \\
    w/o \emph{Crate-Level Bug Isolation}
      & 40.00\% & 53.75\% & 64.38\% & 78.12\% \\
    w/o \emph{Module-Level Bug Isolation}
      & 41.88\% & 57.50\% & 66.25\% & 73.75\% \\
    w/o both isolation components
      & 36.88\% & 56.25\% & 64.38\% & 71.88\% \\
    \midrule
    \multicolumn{5}{l}{\textit{Documentation generation components}} \\
    w/o \emph{Crate-Level Documentation Generation}
      & 37.50\% & 55.62\% & 65.62\% & 77.50\% \\
    w/o \emph{Module-Level Documentation Generation}
      & 41.25\% & 55.62\% & 65.00\% & 76.25\% \\
    w/o \emph{File-Level Documentation Generation}
      & 36.25\% & 54.37\% & 61.25\% & 68.75\% \\
    \midrule
    \multicolumn{5}{l}{\textit{Context management mechanisms}} \\
    w/o \emph{Scoped Candidate Competition}
      & 38.75\% & 54.37\% & 60.62\% & 71.88\% \\
    w/o \emph{Evidence-Guided Code Slicing}
      & 38.75\% & 59.38\% & 68.12\% & 72.50\% \\
    \bottomrule
  \end{tabular}
\end{table}

\paragraph{Experimental Analysis.}

Using the decrease in Accuracy@10 as the comparison criterion, \emph{File-Level Documentation Generation} contributes the most, with a 13.13\% loss when removed, whereas \emph{Crate-Level Bug Isolation} contributes the least, with a 3.76\% loss.  The 10.00\% loss without \emph{Scoped Candidate Competition} also places candidate preservation among the strongest contributors.

\underline{Regarding isolation components}, removing \emph{Module-Level Bug Isolation} reduces Accuracy@10 by 8.13\%, whereas Accuracy@1 falls by only 0.62\%.  Module selection organizes file comparisons within relevant compiler functions, supporting retention of responsible files beyond the first position.  Removing both isolation components increases the Accuracy@10 loss to 10.00\%, supporting their complementary refinement.  The smaller Accuracy@10 loss without crate selection is consistent with the remaining module filter, despite a 6.25\% decrease at Accuracy@3.

\underline{Regarding documentation generation components}, removing \emph{File-Level Documentation Generation} also reduces Accuracy@1 by 6.25\% and Accuracy@5 by 8.75\%.  File descriptions identify concrete checks, analyses, or transformations, distinguishing files with similar names but different responsibilities.  They guide both initial file competition and subsequent relevance analysis, so their absence weakens two successive decisions.  The losses across all cutoffs show that broad descriptions cannot replace this precise account of file functionality.  Removing crate or module documentation causes smaller Accuracy@10 losses of 4.38\% or 5.63\%, respectively.  These summaries support navigation between compiler regions, while file descriptions distinguish concrete implementations within those regions.

\underline{Regarding context management mechanisms}, removing \emph{Scoped Candidate Competition} reduces Accuracy@5 by 9.38\% and Accuracy@10 by 10.00\%, compared with 3.75\% at Accuracy@1.  Comparing files separately within each module limits competition from unrelated compiler functions before global ranking.  The larger losses at broader cutoffs are consistent with poorer retention of alternative bug locations when files from different modules compete together.  Removing \emph{Evidence-Guided Code Slicing} reduces Accuracy@10 by 9.38\%, but Accuracy@3 by only 0.62\% and Accuracy@5 by 1.88\%.  This disproportionate loss is consistent with unrelated source code obscuring additional responsible files in the final ranking.  Slicing instead directs analysis toward computations matching the bug evidence, supporting their inclusion within the first ten positions.

\rqanswer{RQ4}{%
All eight ablated variants achieve lower accuracy than the full approach at every reported metric, supporting the contributions of the evaluated components and mechanisms.  Removing the \emph{File-Level Documentation Generation} component produces the largest Accuracy@10 decrease, while the effects of the other ablations vary across metrics.}

\section{Threats to Validity}
\label{sec:threats}

\paragraph{Internal Validity.}
Internal validity may be affected by baseline adaptation, variation in LLM outputs, and data leakage.  \underline{For baseline adaptation}, we preserve each baseline's core localization procedure, prompt design, and reported settings.  RQ1 and RQ2 use the same dataset, rustc revisions, GPT-5.4 model, and evaluation protocol for all methods.  Adapting the baselines to rustc may still affect their performance.  \underline{For variation in LLM outputs}, RQ1 and RQ3 report means over three independent runs.  This reduces dependence on a single run, although three runs cannot fully characterize variation and the ablation results may remain affected.  \underline{For data leakage}, we exclude fixing PRs from all method inputs and analyze the rustc revision on which each failure is reproduced.  Public issues and fixes may still have appeared in undisclosed model training data.

\paragraph{Construct Validity.}
Construct validity may be affected by ground-truth construction and effectiveness measurement.  \underline{For ground-truth construction}, following prior bug-isolation work~\cite{chen2025locagent,jiang2025issue,xu2025flexfl}, we use faulty compiler files as ground truth.  We derive these files from each fixing PR and exclude tests, documentation, and other non-compiler files.  Supporting edits or refactoring may still introduce files unrelated to the root cause.  \underline{For effectiveness measurement}, following prior bug-isolation work~\cite{chen2025locagent,jiang2025issue,xu2025flexfl}, we evaluate ranked lists at multiple values of $k$.  Accuracy@$k$ records whether any ground-truth file appears among the first $k$ candidates.  Thus, Accuracy@$k$ neither requires identifying all responsible files nor evaluates ranking quality beyond the first $k$ candidates.

\paragraph{External Validity.}
External validity may be affected by bug coverage, rustc evolution, and model coverage.  \underline{For bug coverage}, we screen 160 ICE, completeness, and soundness bugs for artifact availability and reproducibility.  The results may not extend to non-reproducible bugs or other failures such as performance regressions and poor diagnostics.  \underline{For rustc evolution}, HIDE analyzes the revision associated with each bug and generates documentation for that revision.  Major architectural changes may still require adjustments to the hierarchy or query traces.  \underline{For model coverage}, we evaluate three LLMs with unchanged candidate-selection settings.  Other models may differ in accuracy and consistency.

\section{Discussion}
\label{sec:discussion}

\subsection{Extension to Method-Level Bug Isolation}
\label{sec:method-level-extension}

This subsection examines whether HIDE can be extended from file-level to method-level bug isolation.  The extension extracts candidate methods from the suspicious files returned by the \emph{File-Level Bug Isolation} component (Section~\ref{sec:file-isolation}).  The \emph{Scoped Candidate Competition} mechanism selects methods independently within each file before merging the selected candidates.  The \emph{Evidence-Guided Code Slicing} mechanism extracts bug-related code slices for each method and retains the complete body of short methods.  The selected methods are then ranked globally.  We compare the extension with the same three baselines using method-level Accuracy@$k$.  Methods modified by the fixing PR serve as ground truth under the same matching rule as the file-level evaluation.

\begin{table}[t]
  \caption{Method-level bug isolation accuracy on rustc.}
  \label{tab:method-level-results}
  \centering
  \small
  \setlength{\tabcolsep}{5pt}
  \begin{tabular}{@{}lcccc@{}}
    \toprule
    Approach & Accuracy@1 & Accuracy@3 & Accuracy@5 & Accuracy@10 \\
    \midrule
    \textbf{HIDE} & \textbf{32.50\%} & \textbf{49.38\%} & \textbf{58.13\%} & \textbf{60.00\%} \\
    Agentless & \underline{26.25\%} & \underline{39.38\%} & 43.75\% & 48.75\% \\
    SoapFL & 25.62\% & \underline{39.38\%} & \underline{45.62\%} & \underline{56.25\%} \\
    GraphLocator & 18.75\% & 26.25\% & 28.12\% & 28.12\% \\
    \bottomrule
  \end{tabular}
\end{table}

Table~\ref{tab:method-level-results} reports the method-level results on rustc.  HIDE achieves the highest accuracy at every reported metric, reaching 32.50\% Accuracy@1 and 60.00\% Accuracy@10.  These values exceed the strongest baseline results by 6.25\% and 3.75\%, respectively.  Its largest advantage is 12.51\% at Accuracy@5, although accuracy increases by only 1.87\% from Accuracy@5 to Accuracy@10.  The results support extending HIDE to methods but also show that returning more method candidates provides limited additional benefit.

\subsection{Transferability to GCC}
\label{sec:gcc-transferability}

This subsection examines whether HIDE can be transferred from rustc to GCC.  We use the dataset released with CompSCAN~\cite{liu2025isolating}.  It contains 1,026 GCC bug instances with the artifacts required for bug isolation.  We select 200 bugs through stratified random sampling with a fixed seed.  The sample covers the six modules in Table~\ref{tab:gcc-sampling} and preserves their approximate proportions.  These modules contain 1,000 of the 1,026 instances.  The GCC adaptation retains both stages and the two context management mechanisms.  It replaces the rustc query trace with a GCC pass trace that records information about compilation passes.  We evaluate HIDE and the three baselines on the sampled bugs using GPT-5.4 and file-level Accuracy@$k$.

\begingroup
\setlength{\intextsep}{8pt}
\begin{table}[t]
  \caption{Module distribution of GCC candidate bugs and sampled bugs.}
  \label{tab:gcc-sampling}
  \centering
  \small
  \setlength{\tabcolsep}{5pt}
  \begin{tabular}{@{}lrrrr@{}}
    \toprule
    Primary module & \multicolumn{2}{c}{Candidate bugs} & \multicolumn{2}{c}{Sampled bugs} \\
    \cmidrule(lr){2-3} \cmidrule(l){4-5}
    & Count & Percentage & Count & Percentage \\
    \midrule
    \texttt{gcc/root}     & 390   & 39.0\% & 78  & 39.0\% \\
    \texttt{gcc/cp}       & 361   & 36.1\% & 72  & 36.0\% \\
    \texttt{gcc/fortran}  & 124   & 12.4\% & 25  & 12.5\% \\
    \texttt{gcc/config}   & 56    & 5.6\%  & 11  & 5.5\% \\
    \texttt{gcc/c}        & 37    & 3.7\%  & 8   & 4.0\% \\
    \texttt{gcc/c-family} & 32    & 3.2\%  & 6   & 3.0\% \\
    \midrule
    Total                 & 1,000 & 100\%  & 200 & 100\% \\
    \bottomrule
  \end{tabular}
\end{table}
\endgroup

Table~\ref{tab:gcc-results} reports the results on the 200 sampled GCC bugs.  HIDE achieves the highest accuracy at every reported metric, reaching 50.00\% Accuracy@1 and 85.50\% Accuracy@10.  SoapFL is the strongest baseline, with corresponding values of 49.50\% and 80.50\%.  HIDE's advantage over SoapFL increases from 0.50\% at Accuracy@1 to 5.00\% at Accuracy@10.  These results support the transferability of HIDE to GCC under the evaluated setting, although the sample covers only six GCC modules.

\begin{table}[t]
  \caption{File-level bug isolation accuracy on GCC.}
  \label{tab:gcc-results}
  \centering
  \small
  \setlength{\tabcolsep}{5pt}
  \begin{tabular}{@{}lcccc@{}}
    \toprule
    Approach & Accuracy@1 & Accuracy@3 & Accuracy@5 & Accuracy@10 \\
    \midrule
    \textbf{HIDE} & \textbf{50.00\%} & \textbf{71.00\%} & \textbf{79.50\%} & \textbf{85.50\%} \\
    Agentless & 46.50\% & 67.50\% & 71.50\% & 77.00\% \\
    SoapFL & \underline{49.50\%} & \underline{70.00\%} & \underline{76.00\%} & \underline{80.50\%} \\
    GraphLocator & 34.00\% & 40.50\% & 41.00\% & 41.00\% \\
    \bottomrule
  \end{tabular}
\end{table}

\section{Related Work}
\label{sec:related-work}

\subsection{Bug Isolation}
\label{sec:related-bug-isolation}

Bug-isolation research can be broadly divided into general bug isolation and compiler-specific bug isolation.  We review both categories below.

\paragraph{General Bug Isolation.}
General bug-isolation approaches~\cite{xia2025demystifying,li2025llm,yu2025orcaloca,zhang2024autocoderover,kang2024quantitative,xu2025flexfl,qin2025s,widyasari2024demystifying,rafi2025llm4fl,chen2025locagent,jiang2025issue,liu2026graphlocator,ouyang2025repograph} are designed to locate suspicious code across diverse software repositories.  Based on their primary evidence, we organize them into semantics-based, execution-based, and dependency-based approaches and discuss what each category can reveal about a rustc failure and what remains difficult to determine.

Semantics-based approaches~\cite{xia2025demystifying,li2025llm,yu2025orcaloca,zhang2024autocoderover} use descriptions of failing behavior to search for files and functions that implement the affected functionality.  Agentless progressively narrows the search from files to more specific code locations~\cite{xia2025demystifying}.  LLM-BL uses an LLM as a zero-shot ranker to assess whether candidate code implements the functionality described in a bug report~\cite{li2025llm}.  OrcaLoca guides agent exploration by prioritizing search actions, dividing complex actions into smaller steps, assessing candidate code against the behavior described in an issue, and pruning context using code distance~\cite{yu2025orcaloca}.  These methods can use symptoms described in bug reports to identify code worth inspecting, but they are not specifically designed to explain how rustc source files work together to implement compiler functionality.  For example, a file that emits an error message mentioned in a bug report may be ranked highly even when the bug occurs in another file that performs a compiler check, analysis, or transformation.

Execution-based approaches~\cite{kang2024quantitative,xu2025flexfl,qin2025s,widyasari2024demystifying,rafi2025llm4fl} use test execution behavior and coverage information to guide the identification and assessment of suspicious code.  AutoFL allows an LLM to invoke tools for exploring repositories and testing, and to explain the resulting suspicious locations~\cite{kang2024quantitative}.  FlexFL generates candidates using existing bug-isolation techniques and then applies an LLM released as open source to inspect and rerank them~\cite{xu2025flexfl}.  SoapFL integrates test behavior, coverage, static program information, documentation enhancement, and staged LLM review into a structured workflow~\cite{qin2025s}.  Execution information can reduce the search space, but compilation involves many source files with different functions.  Knowing which files execute does not necessarily explain which file performs an incorrect check, analysis, or transformation.  Moreover, reranking selected candidates cannot recover a responsible file that was excluded during candidate selection.

Dependency-based approaches~\cite{chen2025locagent,jiang2025issue,liu2026graphlocator,ouyang2025repograph} trace potential bug causes through code dependencies, often represented as graphs that support searches across connected functions and files.  LocAgent builds a heterogeneous repository graph to support reasoning by an LLM agent~\cite{chen2025locagent}, while CoSIL dynamically constructs, expands, and prunes code graphs during iterative search~\cite{jiang2025issue}.  GraphLocator additionally combines a repository graph with a causal issue graph to connect reported symptoms to possible causes~\cite{liu2026graphlocator}.  These graphs help follow dependencies from one function or file to others, but code connections alone may not explain how files across crates jointly implement compiler functionality or which queries are requested during a failing compilation.  A file closely connected to the code that reports a failure is therefore not necessarily responsible for the bug.

HIDE differs from these general bug-isolation approaches in three respects.  First, the \emph{Layered Documentation Aggregation} stage summarizes what individual files do and how files with similar functions work together, providing documentation at the crate, module, and file levels.  Second, the \emph{Hierarchical Bug Isolation} stage uses this documentation and the bug context to progressively narrow candidates across crates, modules, and files, reducing the number of files requiring detailed analysis.  Third, its \emph{Scoped Candidate Competition} and \emph{Evidence-Guided Code Slicing} mechanisms limit the number of candidates and the amount of source code included in each prompt, respectively.  These designs address distributed compiler functionality, the large source file search space, and context overload in rustc bug isolation.

\paragraph{Compiler-Specific Bug Isolation.}
Compiler-specific bug-isolation approaches~\cite{tu2024llm4cbi,liu2024etem,yang2025odfl,li2026multiconf,liu2025isolating} compare compiler executions to identify the code responsible for a failure.  Based on what they modify to obtain these executions, we organize them into program-based, configuration-based, and step-based approaches.

Program-based approaches~\cite{tu2024llm4cbi,liu2024etem} mutate a program that triggers a compiler failure to generate variants that no longer trigger it, then compare compiler coverage across the original program and these variants.  LLM4CBI uses LLMs to guide program mutation and ranks compiler files using the resulting coverage information~\cite{tu2024llm4cbi}.  This approach requires valid variants that avoid the failure while still exercising enough of the original compilation behavior to make the coverage comparison useful.  Generating such variants can be difficult when small changes to the program cause the compiler to perform different checks or optimizations.

Configuration-based approaches~\cite{yang2025odfl,li2026multiconf} keep the test program unchanged and vary compiler options to obtain executions that do and do not trigger the failure.  ODFL changes individual optimization settings and ranks compiler files using coverage differences across these executions~\cite{yang2025odfl}.  This approach avoids generating new test programs, but it requires options whose settings affect whether the failure occurs.  Its use is therefore limited for bugs that cannot be suppressed by changing the available options.

Step-based approaches~\cite{liu2025isolating} selectively skip compilation steps and observe whether the failure disappears.  CompSCAN uses these observations to identify suspicious steps and rank the compiler code that implements them~\cite{liu2025isolating}.  Applying this strategy requires compilation steps that can be skipped while allowing the remaining compilation to proceed.  In rustc, a query is requested when its result is needed, and its implementation may request other queries.  These computations do not generally form a fixed sequence of steps that can be independently skipped, making the strategy difficult to apply directly.

HIDE does not require changing the test program, varying compiler options, or skipping compilation steps to obtain an execution without the failure.  Instead, the \emph{Hierarchical Bug Isolation} stage uses the original bug context, including the issue description, failure program, compiler log, and query trace, to narrow candidates across crates, modules, and files.  The \emph{Layered Documentation Aggregation} stage supports this search by summarizing what individual files do and how files with similar functions work together.  HIDE can therefore search for responsible files even when it is difficult to make the failure disappear through changes to the program or compilation.

\subsection{Rust Software Engineering}
\label{sec:related-rust-software-engineering}

We review research on Rust software engineering~\cite{cho2024rustsan,kayondo2024metasafe,robati2024study,zhang2024beyond,wang2024rustlantis,yang2024rust,liu2025empirical} in three areas: Rust program safety, Rust software maintenance, and Rust compiler testing.  The following paragraphs describe each area and distinguish its goals from locating responsible source files within rustc.

Rust program safety studies~\cite{cho2024rustsan,kayondo2024metasafe} develop methods to detect or prevent invalid memory accesses in Rust programs.  RustSan adapts AddressSanitizer to Rust and uses program analysis to avoid inserting unnecessary checks~\cite{cho2024rustsan}.  MetaSafe protects the information that smart pointers use to check whether memory accesses are allowed~\cite{kayondo2024metasafe}.  These methods aim to protect Rust programs from memory errors, including those caused by unsafe code.  Their target is the safety of the programs being compiled, whereas HIDE searches for bugs in the compiler implementation.

Rust software maintenance studies~\cite{robati2024study,zhang2024beyond} examine bugs in Rust applications and libraries and the code changes used to repair them.  Robati Shirzad and Lam~\cite{robati2024study} mined Rust repositories to identify recurring changes in bug fixes, including changes involving struct fields and calls to \texttt{clone()}.  Zhang et al.~\cite{zhang2024beyond} studied bugs and fixes in Rust projects and the standard library, classifying their causes and symptoms beyond memory safety issues.  These studies help explain how Rust programs fail and how developers fix them, but they do not identify which rustc source files are responsible when the compiler fails.

Rust compiler testing studies~\cite{wang2024rustlantis,yang2024rust,liu2025empirical} develop methods that generate tests to expose compiler failures or analyze reported bugs to help identify what to test.  Rust-twins mutates Rust programs and generates pairs of macros intended to produce the same results, then compares their expansions to expose compiler errors~\cite{yang2024rust}.  Liu et al.~\cite{liu2025empirical} categorized rustc bugs by their causes, symptoms, affected compilation stages, and test case characteristics.  These efforts help discover compiler bugs and explain which language features or compilation stages need further testing.  They do not, however, produce a ranked list of rustc source files for a given failure.

HIDE complements these studies with a hierarchical LLM-based approach that combines compiler documentation with bug evidence to rank suspicious rustc source files after a failure has been observed.  The \emph{Layered Documentation Aggregation} stage summarizes what individual files do and how files with similar functions work together.  The \emph{Hierarchical Bug Isolation} stage uses these descriptions with the issue description, failure program, compiler log, and query trace to produce a ranked list of suspicious files.  The goal is to help developers decide which compiler files to inspect when repairing a reported bug.

\section{Conclusion}
\label{sec:conclusion}

In this paper, we present HIDE, a hierarchical LLM-based approach that combines compiler documentation with bug evidence to rank suspicious rustc source files.  The \emph{Layered Documentation Aggregation} stage summarizes individual files and how files with similar functions work together.  The \emph{Hierarchical Bug Isolation} stage uses this documentation and the bug context to narrow candidates across crates, modules, and files.  Its \emph{Scoped Candidate Competition} and \emph{Evidence-Guided Code Slicing} mechanisms limit the number of candidates and source code included in each prompt.  Experiments on 160 reproducible rustc bugs show that HIDE outperforms Agentless, SoapFL, and GraphLocator at every reported Accuracy@$k$, achieving 42.08\% Accuracy@1 and 80.42\% Accuracy@10 with GPT-5.4 and the lowest average execution time.  Experiments with three LLMs support its robustness, while all eight ablated variants achieve lower accuracy than full HIDE.  Together, these findings support combining compiler documentation, hierarchical candidate selection, and context management for effective rustc bug isolation.

Future work will consider a broader range of compiler bug types, extend HIDE to additional compilers, and improve its method-level localization.

\bibliographystyle{ACM-Reference-Format}
\bibliography{references}

\end{document}